\documentclass[twocolumn]{autart}

\usepackage{amsmath, amssymb, amsfonts}  % Math stuff
\usepackage[mathscr]{eucal}
\usepackage{cases}

\usepackage{bm}
\usepackage{bbm}
\usepackage{dsfont}
\usepackage{xcolor}
\usepackage{ifpdf}

\usepackage{graphics} % for pdf, bitmapped graphics files
\usepackage{graphicx}           % For including graphics
\graphicspath{{./Images/}}

\usepackage{enumitem}  
\setlist[description]{leftmargin=\parindent}

\usepackage{subfigure}

\usepackage{ifpdf}

\newcommand*\mcapinn[2]{\vcenter{\hbox{$\mathsurround=0pt
  \ifx\displaystyle#1\textstyle\else#1\fi\bigcap$}}}

\newcommand*\mcupinn[2]{\vcenter{\hbox{$\mathsurround=0pt
  \ifx\displaystyle#1\textstyle\else#1\fi\bigcup$}}}

\DeclareFontFamily{OT1}{pzc}{}
\DeclareFontShape{OT1}{pzc}{m}{it}{<-> s * [1.200] pzcmi7t}{}
\DeclareMathAlphabet{\mathpzc}{OT1}{pzc}{m}{it}

\usepackage{theorem}
\newtheorem{theorem}{Theorem}
\newtheorem{definition}{Definition}
\newtheorem{corollary}{Corollary}
\newtheorem{lemma}{Lemma}

\newtheorem{proposition}{Proposition}
\newtheorem{assumption}{Assumption}

\def\qed{\hfill \vrule height 5pt width 5pt depth 0pt \medskip}
\newcommand{\proof}{\noindent {\bf Proof. }}

{\theorembodyfont{\upshape}
	\newtheorem{remark}{Remark}
	\newtheorem{example}{Example}
}

\newcommand{\bfone}{\textbf{1}}

\newcommand{\beq}{\begin{equation}}
\newcommand{\eeq}{\end{equation}}
\newcommand{\beqa}{\begin{eqnarray}}
\newcommand{\eeqa}{\end{eqnarray}}
\newcommand{\beqan}{\begin{eqnarray*}}
\newcommand{\eeqan}{\end{eqnarray*}}

\newcommand{\pde}[2]{ \frac{\partial #1}{\partial #2} }

\newcommand{\bite}{\begin{itemize}}
\newcommand{\eite}{\end{itemize}}
\newcommand{\benu}{\begin{enumerate}}
\newcommand{\eenu}{\end{enumerate}}

\begin{document}

\begin{frontmatter}

 \title{A system-theoretic framework for privacy preservation in continuous-time multiagent dynamics\thanksref{footnoteinfo}}

\thanks[footnoteinfo]{Work supported in part by a grant from the Swedish Research Council (grant n. 2015-04390). A preliminary version of this paper was presented at CDC'19 \cite{Altafini2019Dynamical}.}

\author[primo]{Claudio Altafini}\ead{claudio.altafini@liu.se}

\address[primo]{Division of Automatic Control, Dept. of Electrical Engineering,
Link\"oping University, SE-58183, Link\"oping Sweden.}

% $\mathbbm{1} $ $ \mathds{1}$

                             % keyword list or with the 

%%%%%%%%%%%%%%%%%%%%%%%%%%%%%%%%%
\begin{abstract}
In multiagent dynamical systems, privacy protection corresponds to avoid disclosing the initial states of the agents while accomplishing a distributed task. 
The system-theoretic framework described in this paper for this scope, denoted dynamical privacy, relies on introducing output maps which act as masks, rendering the internal states of an agent indiscernible by the other agents. 
Our output masks are local (i.e., decided independently by each agent), time-varying functions asymptotically converging to the true states.
The resulting masked system is also time-varying, and has the original unmasked system as its limit system.
It is shown that dynamical privacy is not compatible with the existence of equilibria. % in the masked system.
Nevertheless the masked system retains the same convergence properties of the original system: the equilibria of the original systems become attractors for the masked system but loose the stability property.
Application of dynamical privacy to popular examples of multiagent dynamics, such as models of social opinions, average consensus and synchronization, is investigated in detail. 
\end{abstract}

\end{frontmatter}

%%%%%%%%%%%%%%%%%%%%%%%%%%%%%%%%%%%%%%%%%%

\section{Introduction}
Most multiagent systems rely intrinsically on collaboration among agents in order to accomplish a joint task. 
Collaboration however means that exchange of information among the agents cannot be dispensed with. 
If the information is sensitive, then questions like respecting the privacy of the individual agents  naturally rise.
Several approaches exist to address this {\em conundrum} of exchanging information without revealing it.
One approach is called differential privacy \cite{Dwork:2006:DP:2097282.2097284,Dwork:2014:AFD:2693052.2693053} and consists, roughly speaking, in corrupting the information being transmitted with a noise from an appropriate distribution so that an observer accessing the transmitted signals can only reconstruct the original data up to a prespecified precision level.
Another approach relies on cryptography. 
Encrypted messages can be exchanged among the agents in various ways, e.g. through trusted third parties \cite{Lazzeretti6855039}, obfuscation \cite{Ambrosin:2017:OOB:3155100.3137573}, or through distributed cryptography schemes \cite{Ruan19}. 
In these approaches the messages from each agent (corrupted with noise or encrypted) are typically exchanged through a communication graph and hence they are available to the other agents of the network. 
Only the protection mechanism (noise source or cryptographic scheme) is kept private by each agent. 

Both approaches have been recently used for multiagent dynamical systems \cite{Cortes7798915,Hale8031339,Huang:2012:DPI:2381966.2381978,NOZARI2017221,LeNy6606817,Ruan19,Wang7833044}. 
In this case the information to keep private is typically the initial state of the agents.
A problem that is often studied in this context is the consensus problem, because it can be used as a basic building block in many distributed algorithms in database computations, sensor fusion, load balancing, clock synchronization, etc.
Dynamically, a consensus scheme consists of a  stable system in which the final value reached asymptotically is the (weighted) mean of the initial conditions of the agents. 
A privacy protected consensus should render this value available to all agents while not disclosing the initial conditions themselves to the other agents. 
For instance, differentially private average consensus schemes are proposed in \cite{GUPTA20179515,Huang:2012:DPI:2381966.2381978,NOZARI2017221}. 
Clearly the addition of noise impacts also the performances of the consensus algorithm: convergence to the true value might be missing \cite{Huang:2012:DPI:2381966.2381978} or be guaranteed only in expectation \cite{NOZARI2017221}.
% Also the variance of this convergence is subject to tradeoffs \cite{GUPTA20179515}.
Many other variants are possible: for instance in \cite{He2019Consensus,Manitara6669251,Mo7465717,Rezazadeh2018Privacy}, a non-stochastic perturbation is injected at the nodes, with the constraint that the sum (or integral) over time vanishes.
A cryptography-based approach requires instead one or more layers of data encryption technology which must themselves be kept secure and protected \cite{Lazzeretti6855039,Ruan19}.
Other system-oriented approaches to privacy protection in distributed computations appear e.g. in \cite{Alaeddini7963642,Duan7402925,FAROKHI2016254,Kia:RNC3178,Liu2019Dynamical,Monshizadeh19Plausible,Pequito7039593,XUE2014852}.

The aim of this paper is to propose a conceptually different framework for privacy preservation of the initial states of multiagent dynamics, inspired by system-theoretic considerations. 
Our framework is exact, and is developed for continuous-time dynamical systems. It relies on what we call {\em output masks}, i.e., local (in the sense of ``agent-local'', that is, decided and implemented independently by each agent) time-varying transformations of the states to be transmitted to the neighboring nodes, whose functional form and/or numerical parameters are kept hidden to the other agents. 
In the privacy literature, the use of masking maps is widespread. For instance, non-invertible maps are used in homomorphic or semi-homomorphic encryption \cite{Kogiso,FAROKHI201713,Lazzeretti6855039,Ruan19}, as well as in secure wiretap channels \cite{Wiese2016Secure}. 
In the present context, output masks are used to offset the initial condition in a way such that an eavesdropping (curious but not malicious) agent cannot reconstruct it, neither directly nor using a model of the system.
In fact, even when an eavesdropper has knowledge of the vector field used by an agent, reconstruction of the initial state of that agent requires to set up a state observer, which in turn requires to identify the functional form and the numerical parameters of the output mask of the agent.
In the paper this joint ``system identification'' and ``initial state detection'' problem is called {\em discernibility}, and conditions are given that render the initial state indiscernible. 

The approach we follow (offsetting the initial condition) is somewhat related to \cite{He2019Consensus,Mo7465717,Rezazadeh2018Privacy}. However, our use of output masks enables us to carry out a thorough analysis of the dynamical properties of the masked system, which is novel and insightful of the implications of preserving privacy on the dynamics.
When the original unmasked system is globally exponentially stable, or perhaps exponentially stable on ``slices'' of the state space if there is a continuum of equilibria, as in the consensus problem, we show in the paper that under the assumption that no agent has in-neighborhood that covers that of another agent \cite{Rezazadeh2018Privacy,Mo7465717}, the masked multiagent system globally uniformly converges to the same attractor as the unmasked system while guaranteeing the privacy of the initial conditions at any level of precision.

The price to pay for guaranteeing privacy is that the masked system is time-varying and has no fixed points.
However, as long as the output masks are constructed to converge asymptotically to the unmasked state, the masked time-varying system has the original system as its limit system \cite{Artstein1976,ARTSTEIN1977184}. 
When the unmasked system is autonomous, the resulting masked time-varying system is a case of a so-called asymptotically autonomous system \cite{Artstein1976,Markus1956}.

In spite of the indiscernibility of the initial conditions, the asymptotic collapse of the masked dynamics to the original dynamics guarantees that the distributed computation is carried out correctly anyway. 
Clearly, dealing with a distributed computation representable as a converging dynamical system is a key prerequisite of our method, hence we refer to it as {\em dynamical privacy}.

The system-theoretical framework for dynamical privacy developed in this paper is for continuous-time multiagent dynamics. Unlike \cite{Rezazadeh2018Privacy}, where a similar setting is chosen, we do not require the time integrals of the perturbations to be vanishing asymptotically, which gives us more freedom in the choice of the masks and leads to a framework applicable to a broad range of distributed multiagent scenarios.

In the paper we investigate the effect of output masks on three different case studies: a globally exponentially stable nonlinear system, an average consensus problem, and a system of diffusively coupled higher order ODEs achieving pinning synchronization \cite{Chen4232574,Yu-doi:10.1137/100781699,ZHOU2008996}. 
In all three cases a privacy preserving version of the system based on output masks is shown to have the equilibrium point of the unmasked system as unique attractor.
However, as the masked system lacks fixed points, it cannot be stable at the attractor. 
This behavior is designed in purpose.
Think for instance at a situation in which the initial conditions are all in a neighborhood of the (say, globally stable) equilibrium point of the unmasked system. 
If the masked system is stable around that point, its trajectories remain confined in a neighborhood of the equilibrium for all times, leading to an approximate disclosure of the initial states. 
In order to avoid such situations, a masked system cannot preserve neighborhoods of its attractor, or, in other words, the attractor cannot be also an equilibrium point. 
To achieve this, our output masks have to be inhomogeneous in the state variables. 
Such structure is reminiscent of the additive noise used e.g. in differential privacy. 

Technically, to show global attractivity in the masked system (in the complement of the agreement subspace for consensus-like problems), we use Lyapunov arguments. 
The Lyapunov function of the unmasked system is shown to lead to Lyapunov derivatives which are in general sign indefinite, but upper bounded by terms that decay to $0$ as $ t\to \infty $ \cite{MuASJC}. 
The reasoning is fundamentally different from those used in stability analysis of time-varying systems \cite{Aeyels701102,Lee2001,Loria1393135}, but somehow related to constructions used in input-to-state stability \cite{SONTAG1995351} and in the stability analysis of nonlinear systems in presence of additive exponentially decaying disturbances \cite{Sussmann1991Peaking}.
In particular, our masked system has a so-called converging-input converging-state property \cite{Sontag2003RemarkCICS}. 
Boundedness of its trajectories is imposed by choosing Lyapunov functions with globally bounded gradients \cite{Sontag2003Example,Teel2004Examples}. 
The argument is reminiscent of those used in cascade systems \cite{CHAILLET2008519,PANTELEY1998Global,saberi1990global} or in observer-based nonlinear control 
\cite{Arcak2001Observer}.

While the importance of initial conditions is well-known in problems such as average consensus (the final value changes with the initial condition, hence privacy questions are self-evident) in the paper we show that similar privacy issues may arise also in other cases in which the unmasked system is globally exponentially stable. 
In particular we show that in continuous-time Friedkin-Johnsen models of opinion dynamics \cite{PROSKURNIKOV201765}, the value of the equilibrium point is also a function of the initial conditions, because an inhomogeneous term, depending on the initial conditions, is added to an asymptotically stable linear system.
Clearly this is a context in which non-disclosure of the initial states could be of strong relevance. 

The case of pinned synchronization is instead an example of an unmasked system which is time-varying (it depends on the pinning exosystem \cite{Yu-doi:10.1137/100781699,ZHOU2008996}). 
Our privacy protection framework applies also to this case, the only difference being that the limit system of the masked system is itself time-varying.

The rest of the paper is organized as follows: a few preliminary results are outlined in Section~\ref{sec:prelim}, while the dynamical privacy problem and the properties of the output masks are formulated in Section~\ref{sec:prob-form}. In Section~\ref{sec:glob-as-stable} the case of a globally exponentially stable unmasked system (and the related case of Friedkin-Johnsen opinion dynamics model) is discussed. Sections~\ref{sec:av-consensus} and~\ref{sec:synchro} deal with privacy preservation respectively for the average consensus problem and for a pinning synchronization problem.
The proofs of all results are gather in the Appendix. 

In the conference version of this paper, \cite{Altafini2019Dynamical}, only the average consensus problem of Section~\ref{sec:av-consensus} is discussed. The material of Sections~\ref{sec:glob-as-stable} and~\ref{sec:synchro} is presented here for the first time.

%%%%%%%%%%%%%%%%%%%%%%%%%%%%%%%%%%%%%%%%%%

\section{Preliminaries}
\label{sec:prelim}

% {Let $ C^1 $ denote a continuously differentiable function.}
A continuous function $ \alpha \, : \, [0, \infty) \to [ 0, \, \infty) $ is said to belong to class $ \mathcal{K}_\infty $ if it is strictly increasing and $ \alpha(0) =0$. 
Subclasses of $ \mathcal{K}_\infty $ which are homogeneous polynomials of order $ i$ will be denoted $ \mathcal{K}_\infty^i $: $ \alpha(r) = a r^i $ for some constant $ a>0$.
A continuous function $ \zeta \, : \, [0, \infty) \to [ 0, \, \infty) $ is said to belong to class $ \mathcal{L} $ if it is decreasing and $ \lim_{t\to \infty} \zeta(t) =0$. 
In particular, we are interested in $ \mathcal{L} $ functions that are exponentially decreasing: $ \zeta(t) = a e^{-\delta t} $ for some $ a>0$ and $ \delta>0$. We shall denote such subclass $ \mathcal{L}^e \subset  \mathcal{L} $.
A continuous function $ \beta \, : \, [0, \infty) \times [0, \, \infty) \to [ 0, \, \infty) $ is said to belong to class $ \mathcal{KL}_\infty^{i,e} $ if the mapping $ \beta(r, \, t ) $ belongs to class $ \mathcal{K}_\infty^i $ for each fixed $ t $ and to class $ \mathcal{L}^e $ for each fixed $ r$, i.e., $ \beta(r, \, t) = a r^i  e^{-\delta t }$ for some $ a>0$ and $\delta>0$.

Consider
\beq
\dot x = g(t, \, x ), \qquad x(t_o) =x_o 
\label{eq:ode_f(t,x)}
\eeq
where $ g\, : \, \mathbb{R}_+ \times \mathbb{R}^n \to \mathbb{R}^n$ is Lipschitz continuous in $ x$, measurable in $ t$, and such that for each $ x_o \in \mathbb{R}^n $ and each $ t_o \in \mathbb{R}_+ $ the solution of  \eqref{eq:ode_f(t,x)}, $ x(t,  \, x_o ) $, exists in $[0, \, \infty)$.  A point $ x^\ast \in \mathbb{R}^n $ is an equilibrium point of \eqref{eq:ode_f(t,x)} if $ g(t, \, x^\ast ) =0 $ for a.e.\footnote{almost every, i.e., except for at most a set of Lebesgue measure $0$.} $  t \geq t_o$. 

% from Bacciotti p. 
A point $ x^\ast \in \mathbb{R}^n $ is {\em uniformly globally attractive} for \eqref{eq:ode_f(t,x)} if for each $ \nu >0 $ there exists $ T = T(\nu) > 0 $ such that for each solution $ x(t, x_o ) $ of \eqref{eq:ode_f(t,x)} it holds that $ \| x(t, \, x_o ) - x^\ast \| < \nu$ for each $ t > t_o + T $, each $ x_o \in \mathbb{R}^n $ and each $ t_o \geq 0$. 
In particular, if $ x^\ast $ is a uniform global attractor for \eqref{eq:ode_f(t,x)}, then as $ t\to \infty $ all trajectories $ x(t,  x_o) $ converge to $ x^\ast $ %$ \lim_{t\to \infty} x (t+ t_o,  \, x_o ) = x^\ast$ 
uniformly in $ t$ for all $ t_o \geq 0 $ and $ x_o $. 
A point $ x^\ast $ can be attractive  for \eqref{eq:ode_f(t,x)} without being an equilibrium of \eqref{eq:ode_f(t,x)} (we will use this fact extensively in the paper).

Given \eqref{eq:ode_f(t,x)}, denote $ g_s(t, \, x ) $ the translate of $ g(t, \, x)$: $ g_s(t, \, x ) = g(t+s, \, x )$. 
A (possibly time-dependent) system $ \dot x = \tilde g (t, \, x) $ is called a {\em limit system} of \eqref{eq:ode_f(t,x)} if there exists a sequence $ \{ s_k\} $, $ s_k \to \infty $ as $ k\to \infty$, such that $ g_{s_k}(t, \, x ) $ converges to $ \tilde g (t, \, x)$ \cite{Artstein1976}. 
An existence condition for a limit system $ \tilde g(t, \, x) $ is given in Lemma~1 of \cite{Lee2001}: when $ g(t, \, x ) $ is a uniformly continuous and bounded function, then there exist increasing and diverging sequences $ \{ s_k \} $ such that  on compact subsets of $ \mathbb{R}^n$ $ g_{s_k} (t, \, x ) $ converges uniformly to a continuous limit function $ \tilde g(t, \, x) $ on every compact of $ [0, \, \infty)$, as $ k\to \infty$. 
In general the limit system may not be unique nor time-invariant. 
However, when it exists unique, then it must be autonomous \cite{Artstein1976,rouche2012stability} because all translates $ g_{s+s'} (t, \, x ) $ must have themselves a limit system hence the latter cannot depend on time.
The time-varying system \eqref{eq:ode_f(t,x)} is called {\em asymptotically autonomous} in this case.

The $ \omega$-limit set of $ x(t, \, x_o) $, denoted $ \Omega_{x_o} $, consists of all points $ x^\ast $ such that a sequence $ \{ t_k\}$, with $ t_k \to \infty $ when $ k\to\infty $, exists for which $ \lim_{k\to\infty} x(t_k, \, x_o) = x^\ast $. 
For time-varying systems, if a solution is bounded then the corresponding $ \Omega_{x_o} $ is nonempty, compact and approached by $ x(t,  x_o)$. 
However, it need not be invariant. 
Only for limit systems the invariance property may hold, although not necessarily (it may fail even for asymptotically autonomous systems, see \cite{Artstein1976}).

%%%%%%%%%%%%%%%%%%%%%%%%%%%%%%%%%%%%%%%%%%

\section{Problem formulation}
\label{sec:prob-form}
Consider a distributed dynamical system on a graph with $n$ nodes:
\beq \dot x = f(x)  ,\qquad x(0) = x_o ,
\label{eq:model}
\eeq  
where $ x= \begin{bmatrix} x_1 & \ldots & x_n \end{bmatrix}^T \in \mathbb{R}^n $ is a state vector and $ f =[ f_1\, \ldots f_n ]^T  : \, \mathbb{R}^n \to \mathbb{R}^n $ is a Lipschitz continuous vector field. Standing assumptions in this paper are that \eqref{eq:model} possesses a unique solution continuable on $ [0, \, \infty ) $ for all $ x_o \in \mathbb{R}^n $ and that information can be exchanged only between first neighbors on the graph, i.e., 
\beq
\dot x_i = f_i ( x_i, \, x_j, \, j \in \mathcal{N}_i ) , \qquad i=1, \ldots, n
\label{eq:model_i}
\eeq
with $ \mathcal{N}_i $ the in-neighborhood of node $ i$. 
Furthermore, to avoid trivial situations, we impose that $ \mathcal{N}_i $ is the ``essential neighborhood'' of agent $i$ \cite{Mo7465717}, i.e., 
\beq
\begin{split}
f_i(x_i, \, x_j, \, j \in \tilde{\mathcal{N}}_i ) \neq  f_i (x_i, \, x_j, \, j \in \mathcal{N}_i ) &\quad \forall \tilde{\mathcal{N}}_i \subsetneq  \mathcal{N}_i, \\
& \forall i=1\ldots, n.
\end{split}
\label{eq:non-trivial-neigh}
\eeq

We are interested in cases in which the system \eqref{eq:model_i} has a globally exponentially  stable equilibrium point, i.e., $ \lim_{t\to \infty } x(t) = x^\ast $ for all $ x_o$, but also in cases in which the presence of a conservation law (as in the consensus problem) leads to exponential stability on some submanifold depending on the initial conditions, i.e., $ \lim_{t\to \infty} x(t) = x^\ast(x_o) $.

The {\em privacy preservation problem} consists in using a system like \eqref{eq:model} to perform the computation of $ x^\ast $ in a distributed manner, while avoiding to divulgate the initial condition $ x_o $ to the other nodes. 
Clearly this cannot be achieved directly on the system \eqref{eq:model} which is based on exchanging the values $ x_i $ between the nodes. 
It can however be achieved if we insert a mask on the value $ x(t) $ which preserves convergence to $ x^\ast$, at least asymptotically. 
The masks we propose in this paper have the form of time-varying output maps.

%--------------------------------

\subsection{Output masks}

Consider a continuously differentiable time-varying output map
\beq
\begin{split}
h \, : \, \mathbb{R}_+ \times \mathbb{R}^n \times \mathbb{R}^m & \to  \mathbb{R}^n \\
(t, \, x, \, \pi ) & \mapsto  y(t) = h(t, x(t) , \pi) 
\end{split}
\label{eq:output1}
\eeq
where $ y = \begin{bmatrix} y_1 & \ldots & y_n \end{bmatrix}^T \in \mathbb{R}^n $ is an output vector of the same size as $x$, and $ \pi\in \mathbb{R}^{m} $ is a vector of parameters splittable into $n$ subvectors (not necessarily of the same dimension), one for each node of the network: $ \pi =\{ \pi_1,  \ldots , \pi_n \} $.

In the following we refer to $h(t, x(t) , \pi)$ as an {\em output mask} and to $ y $ as a {\em masked output}.
The state $ x$ of the system is first masked into $y$ and then sent to the first out-neighbors on the graph.
The original system \eqref{eq:model} can therefore be modified into the following {\em masked system:}
\begin{subequations}
\begin{align}
% \begin{cases}
\dot x & = f(y)  \label{eq:model_xy_a}\\ 
y & = h(t, x, \pi) .\label{eq:model_xy_b}
% \end{cases}
\end{align}
\label{eq:model_xy}
\end{subequations}
Denote $ y(t, x_o) $, of components $ y_i(t, x_{o,i})$, $ i=1, \ldots, n$,  the output trajectory of \eqref{eq:model_xy} from the initial state $ x_o$, of components $ x_{o,i}$.
We assume in what follows that the vector field $ f(\cdot) $ is publicly known (i.e., each agent knows the shape of the functions $ f_1(\cdot), \ldots, f_n (\cdot) $) and that each node knows the output trajectories $ y_i(t, x_{o,i})$ of its in-neighbors. The state $ x$ and the output mask $ h(t, x, \pi) $ (functional form plus values of the parameters $ \pi$) are instead private to each agent, as explained more in detail next.

\begin{definition}
\label{def:local_mask}
A $ C^1 $ output map $h$ is said a {\em local mask} if it has components that are local, i.e.,
\benu
\item[P1:] $ h_i(t, x, \pi) =  h_i(t, x_i, \pi_i ) \qquad i =1, \ldots, n$.
\eenu
\end{definition}
The property of locality guarantees that the output map $ h_i $ can be independently decided by node $i$.
Both the functional form chosen for $ h_i(\cdot)$ and the numerical value of the parameters $ \pi_i $ can therefore remain hidden to the other agents.

The output mask needs also to avoid mapping neighborhoods of a point $ x^\ast $ of \eqref{eq:model} (typically an equilibrium point) into themselves.
For that, we introduce the following definition.

\begin{definition}
A $ C^1 $ output map $ h $ is said to {\em preserve neighborhoods} of a point $ x^\ast $ if, for all small $ \epsilon >0$, $ \|x_o- x^\ast \| <\epsilon \;\; \Longrightarrow \;\;  \|h (0, x_o, \pi)- x^\ast \| < \epsilon$. It is said {\em not to preserve neighborhoods} otherwise.
\end{definition}

These notions are used in the following definition.
\begin{definition}
\label{def:privacy_mask}
A $ C^1 $ output map $h$ is said a {\em privacy mask} if it is a local mask and in addition
\benu
\item[P2:] \label{def:mask1}  $ h_i(0, x_i, \pi_i) \neq x_i $ $ \forall \; x_i \in \mathbb{R}^n$, $ i=1, \ldots, n$;
% \item[P3:] \label{def:mask2}  $ h(t,x, \pi) $ guarantees indiscernibility of the initial conditions;
\item[P3:] \label{def:mask3}  $ h(t,x, \pi) $ does not preserve neighborhoods of any $ x \in \mathbb{R}^n$;
\item[P4:] \label{def:mask4} $  h_i(t, x_i, \pi_i) $ is strictly increasing in $ x_i$ for each fixed $ t$ and $ \pi_i$, $ i=1, \ldots, n$.
\eenu
\end{definition}
Property P2 means that $ h_i (\cdot) $ has no fixed points.
Property P4 %\ref{def:mask4}
resembles a definition of $ \mathcal{K}_\infty $ function, but it is in fact more general: $ x=0 $ is not a fixed point of $ h $ for any finite $ t$, and $ h$ need not be nonnegative in $ x$.
Monotonicity of $ h$ in $ x $ (for each fixed $ \pi$) follows from Property P4 combined with P1. It implies that $ h $ is a bijection in $ x $ for each fixed $ t$ and $ \pi$, although one that does not preserve the origin. 
This is meant to avoid that the output mask introduces undesired behavior in the system, like spurious equilibrium points.

In many cases, it will be necessary to impose that the privacy mask converges asymptotically to the true state, i.e., that the perturbation induced by the mask is vanishing.
\begin{definition}
\label{def:vanishing_privacy_mask}
The output map $h$ is said a {\em vanishing privacy mask} if it is a privacy mask and in addition 
\benu
 \item[P5:] \label{def:mask5}$ | h_i(t, x_i, \pi_i ) - x_i | $ is decreasing in $ t$  for each fixed $ x_i$ and $ \pi_i$, and $ \lim_{t\to \infty } h_i(t, x_i, \pi_i ) = x_i $ for each fixed $ \pi_i$, $ i=1, \ldots, n$. % [convergence to the state]
 \eenu
\end{definition}

The difference between the true initial condition $ x_{o,i} $ and the masked output $ h_i(0, x_{o,i}, \pi_i ) $ can be used to quantify the level of privacy for agent $i$. More formally, if $ h_i (\cdot) $ is a privacy mask for agent $i$, we denote $ \rho_i (x_{o,i} ) = | h_i(0, x_{o,i}, \pi_i )  - x_{o,i} | $ the privacy metric of agent $i$ relative to the initial condition $ x_{o,i}$, and $ \rho(x_o) = \min_{i=1, \ldots, n} \rho_i(x_{o,i} ) $ the privacy metric of the system relative to the initial condition $ x_o $. 

%-----------------------------------

\subsection{Examples of output masks}
The following are examples of output masks.
\paragraph*{Linear mask}
\beq
h_i(t, x_i, \pi_i ) = (1 + \phi_i e^{-\sigma_i t } ) x_i , \qquad \phi_i \geq   0, \quad \sigma_i > 0 
\label{eq:ex-linear-mask}
\eeq
(i.e., $ \pi_i = \{ \phi_i, \, \sigma_i \}$). This local vanishing mask is not a proper privacy mask since $ h_i(0, 0, \pi_i ) =0 $ i.e. the origin is not masked. Notice that all homogeneous maps have this problem (and they fail to escape neighborhoods of $ x_i $).
\paragraph*{Additive mask} 
\beq
h_i(t, x_i, \pi_i ) = x_i + \gamma_i e^{-\delta_i t }, \qquad \delta_i >0 , \quad \gamma_i \neq 0 
\label{eq:ex-additive-mask}
\eeq
(i.e., $ \pi_i = \{ \delta_i, \, \gamma_i \}$) is a vanishing privacy mask.

\paragraph*{Affine mask}  
\beq
h_i(t, x_i, \pi_i ) =c_i (x_i + \gamma_i e^{-\delta_i t }), \quad c_i > 1, \quad \delta_i>0, \quad  \gamma_i\neq 0 
\label{eq:ex-affine-mask}
\eeq
(i.e., $ \pi_i = \{ c_i, \, \delta_i, \, \gamma_i \}$) is also a privacy mask.
Since $ \lim_{t\to\infty} h_i (t, \, x_i, \, \pi_i ) = c_i x_i $, it is however not vanishing.
\paragraph*{Vanishing affine mask}
\beq
\begin{split}
h_i(t, x_i, \pi_i ) = & (1 + \phi_i e^{-\sigma_i t } ) (x_i + \gamma_i e^{-\delta_i t }), \\
&  \phi_i > 0, \quad \sigma_i >0, \quad \delta_i>0, \quad  \gamma_i\neq 0 
\end{split}
\label{eq:van-aff-mask-scalar}
\eeq
(i.e., $ \pi_i = \{ \phi_i, \, \sigma_i , \, \delta_i, \, \gamma_i \}$). This privacy mask is also vanishing.
% \end{example}
Notice that in vector form, assuming all nodes adopt it, the vanishing affine mask can be expressed as
\beq
h(t, \, x, \pi)  = (I + \Phi e^{-\Sigma t} ) (x + e^{-\Delta t } \gamma)
\label{eq:output-mask-affine-II}
\eeq
where $ \Phi = {\rm diag}( \phi_1, \ldots, \phi_n ) $, $ \Sigma= {\rm diag}( \sigma_1, \ldots, \sigma_n ) $, $ \Delta = {\rm diag}( \delta_1, \ldots, \delta_n ) $, and $ \gamma = \begin{bmatrix} \gamma_1 & \ldots &  \gamma_n \end{bmatrix}^T $.

The following proposition shows that for these masks the level of privacy (as defined by the metric $ \rho$), can be made arbitrary if each agent chooses $ \pi_i $ in an appropriate way.

\begin{proposition}
\label{prop:metric}
Given $ \lambda>0 $, for each of the privacy masks \eqref{eq:ex-additive-mask}, \eqref{eq:ex-affine-mask} and \eqref{eq:van-aff-mask-scalar} the parameters $ \pi_i$ can be chosen locally by each agent $ i$, $ i =1, \ldots, n $, so that $ \rho(x_o ) > \lambda $ for any $ x_o$. 
\end{proposition}
By definition of $ \rho$, the choice of parameters $ \pi_i $ in Proposition~\ref{prop:metric} is compatible with the privacy of the maps $ h_i(\cdot)$ (each agents knows its own $ x_{o,i}$, hence can compute $ \rho_i (x_{o,i} ) $ without disclosing $ x_{o,i}$).

%---------------------------------------------------------------------

\subsection{Dynamically private systems}

Consider the system \eqref{eq:model_xy}, rewritten here for convenience in components ($i =1, \ldots, n$):
\begin{subequations}
\begin{align}
\dot x_i & = f_i(y_i, \, y_k , \, k \in \mathcal{N}_i  ) , \qquad  x_i(0) = x_{o,i} \label{eq:model_xyi_a} \\
y_i & = h_i(t, \, x_i, \, \pi_i ) . \label{eq:model_xyi_b}
\end{align}
\label{eq:model_xyi}
\end{subequations}
We would like to understand when an eavesdropping agent $j$ can violate the privacy of agent $ i$, estimating its initial condition $ x_{o,i}$. 
Recall that we are assuming that the agent $ j$ knows:
\benu
\item[K1:] \label{list_known2} the form of the vector field $ f_i (\cdot ) $,
\item[K2:] \label{list_known2} the output trajectories of its incoming neighborhood: $ y_k(t, x_{o,k}) $, $ k \in \{ \mathcal{N}_j \cup \{ j \} \} $, $ t \in [t_o, \, \infty)$,
\eenu
while instead the following are unknown for agent $j$:
\benu
\addtocounter{enumi}{2}
\item[U1:] \label{list_unknown2} the form of the output mask $ h_i (\cdot) $ and the numerical values of the parameters $ \pi_i$,
\item[U2:] the output trajectories not in its incoming neighborhood: $ y_k(t, x_{o,k}) $, $ k \notin \{ \mathcal{N}_j \cup \{ j \} \} $.
\eenu

Because of item~U1 above, the problem of estimating $ x_{o,i}$ from the output in \eqref{eq:model_xyi} cannot be cast as  a state observability problem, but rather it has to be treated as a joint system identification + observability problem. To characterize this unusual situation we introduce a new concept, discernibility. 

\begin{definition}
The initial condition of agent $ i$, $ x_{o,i} $, is said {\em discernible for agent $ j$} ($j\neq i$) if agent $ j$ can estimate $ x_{o,i}$  from the knowledge of K1 and K2.
It is said {\em indiscernible for agent $j$} otherwise. 
An initial condition $ x_o$ is said {\em indiscernible} if all of its components  $ x_{o,i}$ are indiscernible for all agents $ j \in \{ 1, \ldots, n \} \setminus \{ i\}$.
\end{definition}
Indiscernibility refers to the impossibility to solve the joint identification+observation problem of estimating $ x_{o,i}$ in \eqref{eq:model_xyi}. It can be imposed using the properties of a privacy mask together with the following Assumption~\ref{ass1} (see  \cite{Rezazadeh2018Privacy}  and \cite{Mo7465717}, Corollary 1).
\begin{assumption}
\label{ass1} {\rm (No completely covering neighborhoods)} 
The system \eqref{eq:model_xy} is such that $ \{ \mathcal{N}_i\cup \{ i \} \} \nsubseteq \{ \mathcal{N}_j \cup \{ j \}\} $, $ \forall \; i,\, j =1, \ldots, n$, $ i \neq j$. 
\end{assumption}
Assumption~\ref{ass1} guarantees that no node has complete information of what is going on at the other nodes.
This is a condition on the topology of the graph, and therefore a {\em system} property, rather than simply a property of well-conceived output maps.

Combining indiscernibility with privacy of the output masks, we can formulate the following definition.
\begin{definition}
\label{def:priv-sys}
The system \eqref{eq:model_xy} is called a {\em dynamically private} version of  \eqref{eq:model} if 
\benu
\item $ h $ is a privacy mask;
\item the solution of  \eqref{eq:model_xy} exists unique in $ [0, \, \infty ) $ and is bounded $ \forall \; x_o \in \mathbb{R}^n$;
\item $ \lim_{t\to \infty} y(t) = \lim_{t\to\infty} x(t) $;
% \item For each $ i=1,\ldots,n$, the integral $ \int_0^\infty f_i(y) dt $ cannot be estimated by an agent $ j \neq i$.
\item indiscernibility of the initial condition is guaranteed.
% \item an external observer cannot reconstruct the initial condition $ x(0) $ from the observation of $ y(t)$. 
\eenu
\end{definition}

The next proposition relates indiscernibility to Assumption~\ref{ass1}.
\begin{proposition}
\label{prop:ass1:indiscern}
If the system \eqref{eq:model_xy} satisfies conditions~1-3 of Definition~\ref{def:priv-sys} and Assumption~\ref{ass1}, then it is a dynamically private version of  \eqref{eq:model}.
\end{proposition}

\begin{remark}
From Proposition~\ref{prop:metric}, when a system is dynamically private with any of \eqref{eq:ex-additive-mask}, \eqref{eq:ex-affine-mask} and \eqref{eq:van-aff-mask-scalar} as mask, then privacy can be made to hold at an arbitrary level of precision, i.e., given $ \lambda>0 $, $ \rho(x_o) >\lambda $ can be guaranteed for any $x_o$ only though local choices of the parameters $ \pi_i $ for each agent.
\end{remark}

The privacy property P3 of $ h(\cdot ) $ suggests that in a dynamically private system we cannot have equilibrium points and therefore we cannot talk about stability (of equilibria), while convergence of $ y(t)$ to $ x(t)$ suggests that as long as $ f(\cdot ) $ is autonomous, a dynamically private system is asymptotically autonomous with the unmasked system as limit system. 
This can be shown to be always true if the output mask is vanishing. 

\begin{proposition}
\label{prop:no-equil}
If \eqref{eq:model_xy} is a dynamically private version of \eqref{eq:model}, then it cannot have equilibrium points.
Furthermore, if $ h (\cdot ) $ is a vanishing privacy mask, then the system \eqref{eq:model_xy} is asymptotically autonomous with limit system \eqref{eq:model}.
\end{proposition}

The ``vanishing'' attribute of the second part of Proposition~\ref{prop:no-equil} is sufficient but not necessary. As we will see below, when \eqref{eq:model} is globally exponentially stable, the condition that the output mask must be vanishing can be dispensed with.

%%%%%%%%%%%%%%%%%%%%%%%%%%%%%%%%%%%%%%%%%%%%%

\section{Dynamical privacy in globally exponentially stable systems}
\label{sec:glob-as-stable}
{In this section we restrict ourselves to unmasked systems \eqref{eq:model} having a globally exponentially stable equilibrium point. Under Assumption~\ref{ass1}, any privacy mask (not necessarily vanishing) can guarantee privacy of the initial conditions. We will only show the simplest case of affine mask. 
Since we rely on standard converse Lyapunov theorems, we also request \eqref{eq:model} to be globally Lipschitz. }

\begin{theorem}
\label{thm:glob-as-stab}
Consider the system \eqref{eq:model} with $ f\, : \, \mathbb{R}^n \to \mathbb{R}^n$ {globally} Lipschitz continuous, $ f(0) =0$, and the masked system \eqref{eq:model_xy} with the affine mask \beq
h(t,x,p)= C(x + e^{-\Delta t} \gamma),
\label{eq:glob-as-st-output-mask}
\eeq 
$ C= {\rm diag}(c_1, \ldots, c_n )$, $ c_i > 1 $, $ \Delta = {\rm diag} (\delta_1, \ldots, \delta_n)$, $ \delta_i > 0$, and $ \gamma = \begin{bmatrix} \gamma_1 & \ldots &  \gamma_n \end{bmatrix}^T $, $ \gamma_i\neq 0 $. 
If Assumption~\ref{ass1} holds and the equilibrium $ x^\ast =0 $ is globally {exponentially} stable for \eqref{eq:model}, then $ x^\ast =0 $ is uniformly globally attractive for the masked system \eqref{eq:model_xy}. Furthermore, \eqref{eq:model_xy} is a dynamically private version of \eqref{eq:model}.
\end{theorem}

\begin{remark}
Even if  \eqref{eq:model} has $ x^\ast =0 $ as equilibrium point, the masked system  \eqref{eq:model_xy} does not, as can be seen from the expression \eqref{eq:doty} in the proof of Theorem~\ref{thm:glob-as-stab}.
This follows from the inhomogeneity of the output mask.
Since $ x^\ast =0 $ is not stationary, we cannot talk about stability of its neighborhoods. 
Nevertheless, $ x^\ast $ remains an attractor for all trajectories of the system.
\end{remark}

The following corollary states that the dynamically private system is asymptotically autonomous with $ \omega$-limit set identical to that of the corresponding unmasked system.
\begin{corollary}
\label{cor:as-auton}
Under the assumptions of Theorem~\ref{thm:glob-as-stab}, the system \eqref{eq:model_xy} with the output mask \eqref{eq:glob-as-st-output-mask} is asymptotically autonomous with limit system 
\beq
 \dot x = f(C^{-1} x ) .
\label{eq:glob-as-st-limit}
\eeq
The $\omega $-limit set of each trajectory of \eqref{eq:model_xy} is given by $\{ 0 \} $ for each $ x_o \in \mathbb{R}^n$. 
\end{corollary}

Notice that since the affine mask \eqref{eq:glob-as-st-output-mask} is not vanishing, \eqref{eq:glob-as-st-limit} differs from \eqref{eq:model} (yet $ x^\ast $ is the same).

\begin{remark}
The result of Theorem~\ref{thm:glob-as-stab} can be rephrased as a converging-input converging-state property \cite{Sontag2003RemarkCICS}: under the assumption of $ f$ (locally) Lipschitz continuous and $ x^\ast $ globally asymptotically stable, boundedness of the trajectories %of \eqref{eq:doty} 
is enough to guarantee that %$ y(t) \to 0 $ (and hence $ x(t) \to 0 $)
$ x(t) \to 0 $ as $ t\to \infty$. 
However, guaranteeing boundedness is a nontrivial task: a globally asymptotically stable system can be destabilized by an additive perturbation which is arbitrarily small in $ \mathcal{L}_1 $ norm \cite{Sontag2003Example}.
Similarly, a globally exponentially stable system with linear sector growth (as opposed to global Lipschitzianity) can be destabilized by arbitrarily small additive exponentially decaying disturbances \cite{Teel2004Examples,Astolfi2007Remark}.
The assumptions made in Theorem~\ref{thm:glob-as-stab} imply the boundedness of the gradient of the Lyapunov function, which in turn guarantees boundedness of the solutions.
\end{remark}

\begin{example}
\label{ex:globally-exp-stable}
Consider the following interconnected system with saturated nonlinearities
\beq
\dot x = -x + \kappa A \psi (x)
\label{eq:interconn-example}
\eeq
where the off-diagonal matrix $ A \geq 0\; $  is a weighted adjacency matrix of spectral radius $ \rho(A)>0 $ describing the interactions among the agents and satisfying Assumption~\ref{ass1}, $ \kappa> 0 $ is a scalar coefficient, and $ \psi(x) = \begin{bmatrix} \psi_1(x_1) & \ldots & \psi_n(x_n) \end{bmatrix}^T $, $ \psi_i(x_i) = \tanh (x_i) $, is a vector of saturated sigmoidal functions depending only on the state of the sending node $ x_i $. 
The system \eqref{eq:interconn-example} is used e.g. in \cite{FoAl2018} to describe collective distributed decision-making systems. 
If we impose the condition $ \kappa< \frac{1}{\rho(A)} $, then $ x^\ast =0 $ is a globally exponentially stable equilibrium point for \eqref{eq:interconn-example}. In fact, in this case a simple quadratic Lyapunov function $ V= \frac{1}{2} \| x \|^2 $ leads to
\[
\dot V = - x^T x + \kappa x^T A \psi(x) \leq x^T ( -I + \kappa A ) x<0
\]
because $ \psi_i (x_i) $ obeys to the sector inequality $ 0 \leq \psi_i(x_i) x_i \leq 1 $.
Since the system is globally Lipschitz,  Theorem~\ref{thm:glob-as-stab} is applicable to it if we choose an output mask like \eqref{eq:glob-as-st-output-mask}. Simulations for $ n= 100 $ are shown in Fig.~\ref{fig:exp-stable} for a privacy measure of $ \lambda=1 $. 
Notice in panel (c) how the gap between $ x_{o,i} $ and $ y_{o,i} $ induced by this privacy level is clearly visible.
% This privacy requirements is clearly visible in the distribution of $ y(0)$, see panel (c).
The initial conditions obey to $ \rho_i (x_{o,i}) = |y_i(0) - x_i(0)|\geq 1$, but  $ | y_i(t) - x_i(t) | $ necessarily decreases as t grows, and converges to 0 as $ t$ diverges, see panel (d).  
\begin{figure*}[htb]
\begin{center}
\subfigure[]{
\includegraphics[angle=0, trim=1cm 7.5cm 9cm 1cm, clip=true, width=4.2cm]{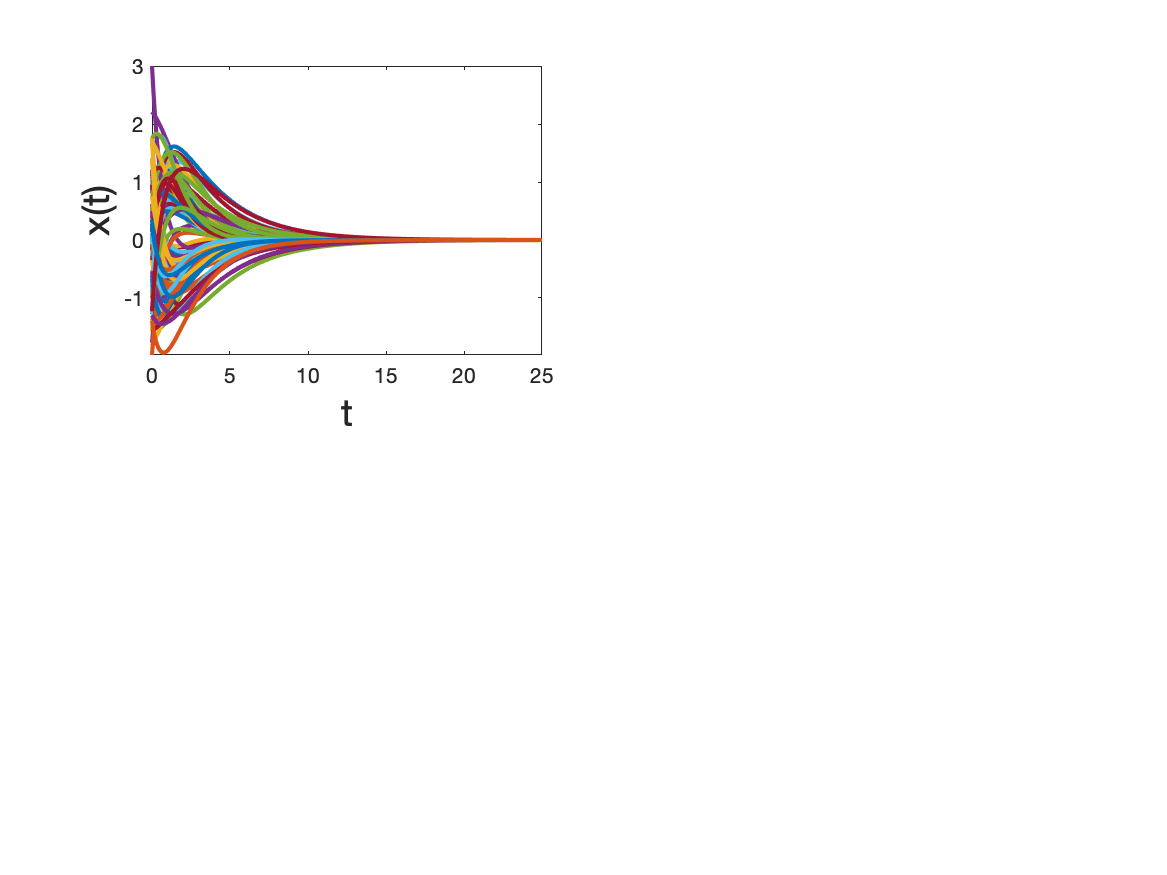}} 
\subfigure[]{
\includegraphics[angle=0, trim=1cm 7.5cm 9cm 1cm, clip=true, width=4.2cm]{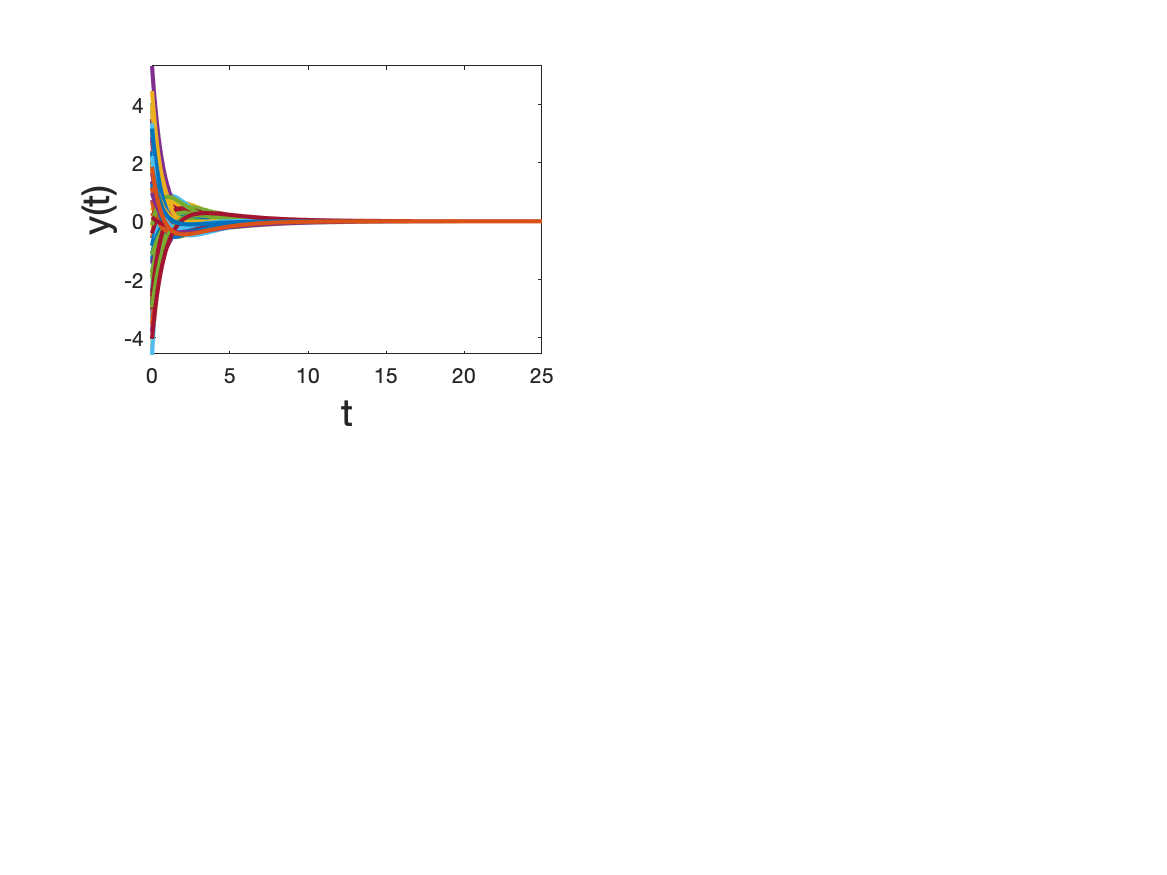}}
\subfigure[]{
\includegraphics[angle=0, trim=1cm 7.5cm 9cm 1cm, clip=true, width=4.2cm]{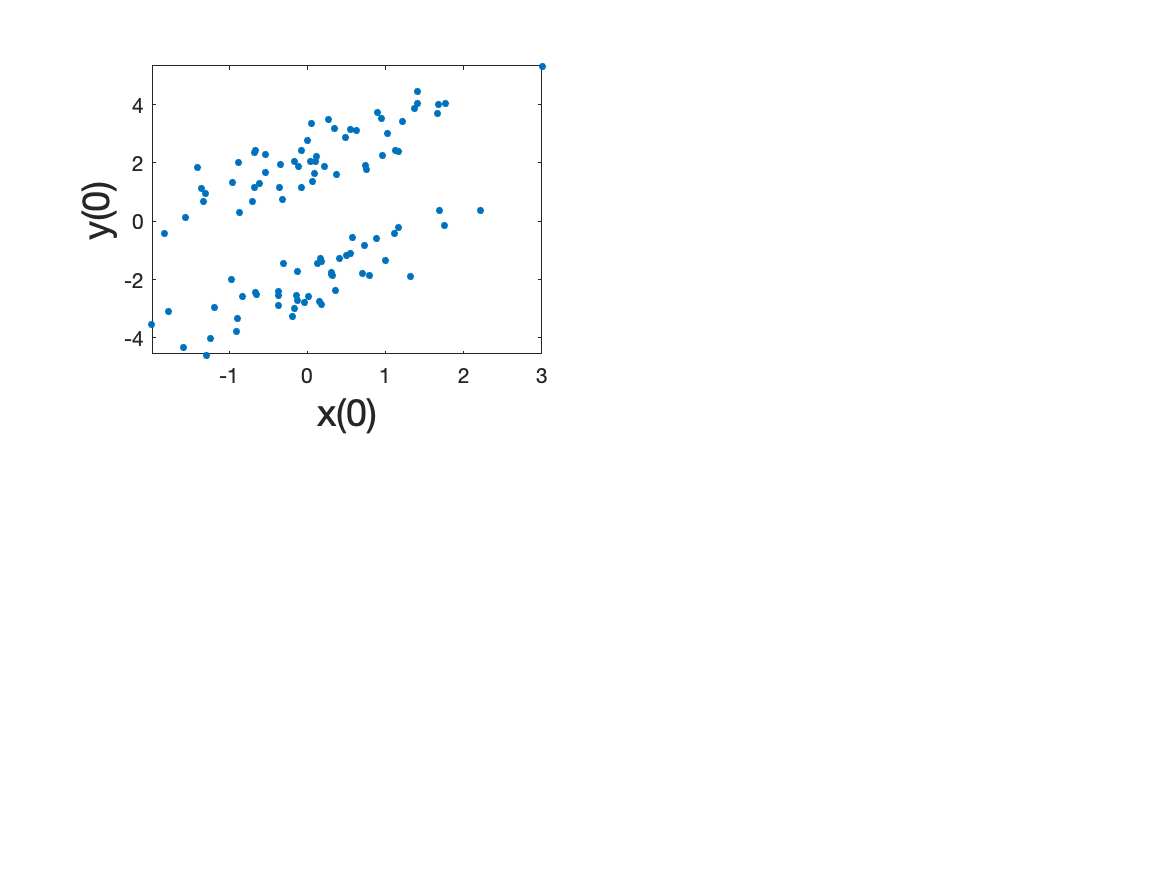}}
\subfigure[]{
\includegraphics[angle=0, trim=1cm 7.5cm 9cm 1cm, clip=true, width=4.2cm]{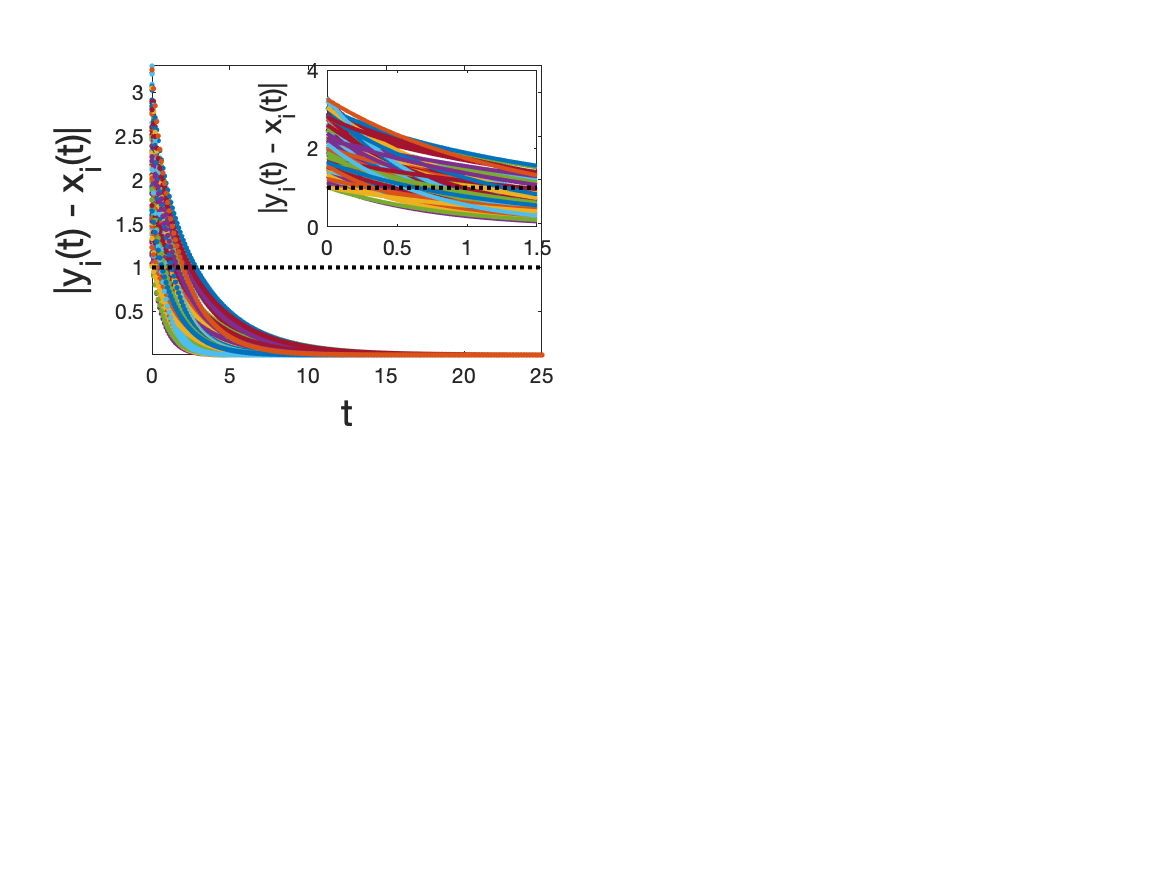}}
\caption{Privacy-preserving globally exponentially stable system of Example~\ref{ex:globally-exp-stable}. (a): private state $x(t)$; (b): masked output $y(t)$; (c): initial conditions $ x(0) $ vs. $ y(0)$; (d): $ |y_i(t) - x_i(t)|$. The black dotted line in (d) represents $ \lambda $. The inset of panel (d) is a zoom in of the initial part.}
\label{fig:exp-stable}
\end{center}
\end{figure*}
\end{example}
In Example~\ref{ex:globally-exp-stable} global exponential stability implies that the initial conditions are forgotten asymptotically. In these cases privacy protection might be considered less critical than when the equilibrium point is itself a function of the initial state, as it happens in the next sections.

%---------------------------------------------------------------------------------------

\subsection{Application to continuous-time Friedkin-Johnsen model}
Let us consider a continuous-time Friedkin-Johnsen model (also known as Taylor model, see \cite{PROSKURNIKOV201765})
\beq
\dot x = - (L + \Theta ) x + \Theta x_o , \qquad x(0) =x_o,
\label{eq:FJ_cont}
\eeq
where $ L $ is an irreducible Laplacian matrix, and $ \Theta = {\rm diag}(\theta_1, \ldots, \theta_n ) $, $ \theta_i \in [0, \, 1 ] $, is a diagonal matrix of so-called susceptibilities, i.e., tendencies of the $ i$-th agent to remain attached to its own initial opinion $ x_{o_i}$. 
The behavior of the system \eqref{eq:FJ_cont} is analyzed in \cite{PROSKURNIKOV201765}: when $ L $ is irreducible and some $ \theta_i \neq 0 $, it has a single equilibrium point $ x^\ast = (L+\Theta)^{-1} \Theta x_o %\in  {\rm co} (x_o, \bfone^T x_o /n \bfone )
$ which is asymptotically stable for a solution starting in $ x_o $. 
The system reduces to the usual consensus problem when $ \theta_i =0 $ $ \forall \, i $ (see Section~\ref{sec:av-consensus}).
Notice how in the affine model \eqref{eq:FJ_cont}, the initial opinions (initial condition of the system) enter also in the vector field at time $ t$. 
Hence protecting the privacy of the agents in \eqref{eq:FJ_cont} requires a `double mask', i.e., one needs to replace both $ x (t) $ and $ x_o $ with suitably masked versions $ y(t)$ and $ y_o = y(0)$ (since $ y_o $ is transmitted to the neighboring agents, it can be memorized and used whenever needed). 

Denoting $ z = x - x^\ast = x - (L+\Theta)^{-1} \Theta x_o $, then \eqref{eq:FJ_cont} is expressed in $z $ as the linear system
\beq
\dot z = -(L + \Theta) z ,
\label{eq:FJ_cont-z}
\eeq
which has $ z^\ast =0 $ as globally asymptotically (and hence exponentially) stable equilibrium point, meaning that Theorem~\ref{thm:glob-as-stab} is applicable.
In the original $ x$ basis, a consequence of inhomogeneity of \eqref{eq:FJ_cont} is that the attractor $ x^\ast $ is a function of the initial condition $ x_o $, and it moves with it: $ x^\ast = x^\ast (x_o)$. 
To talk rigorously about global asymptotic stability, we should use \eqref{eq:FJ_cont-z} in $ z$-coordinates.
However, for homogeneity of presentation, the next theorem is still formulated in terms of $ x $ and $ y $ variables, and global asymptotic stability / attractivity is referred to the ``moving'' point\footnote{Unlike for the consensus problem which we will study in Section~\ref{sec:av-consensus}, in this case there is no easy way to describe the orthogonal complement of the space in which $ x^\ast(x_o)$ moves.} $ x^\ast(x_o)$. 
Another consequence of the inhomogeneous structure of \eqref{eq:FJ_cont} is that novanishing affine privacy masks like the one used in Theorem~\ref{thm:glob-as-stab} cannot be used. To obtain convergence to the correct $ x^\ast (x_o)$ we need to use a vanishing privacy mask. 

\begin{theorem}
\label{thm:FJ}
If Assumption~\ref{ass1} holds, the  masked system
\beq
\begin{split}
\dot x & =  (-L - \Theta ) y + \Theta y_o \\
y & = h(t, \,  x, \, \pi ) = \left(I + \Phi e^{-\Sigma t } \right)  \left( x + e^{-\Delta t} \gamma \right) 
\end{split}
\label{eq:FJ_cont_private}
\eeq
where $ y_o  = h(t, \,  x_o, \, \pi ) $ and $ \Theta\neq 0$, is a dynamically private version of \eqref{eq:FJ_cont}.
If $ x^\ast (x_o) = (L+\Theta)^{-1} \Theta x_o $ is the globally asymptotically stable equilibrium point of \eqref{eq:FJ_cont}, then $ x^\ast (x_o) $ is a globally uniform attractor of \eqref{eq:FJ_cont_private}.
\end{theorem}

\begin{corollary}
\label{cor:FJ-autonomous}
The masked system \eqref{eq:FJ_cont_private} is asymptotically autonomous with \eqref{eq:FJ_cont} as limit system. 
The $ \omega$-limit set of \eqref{eq:FJ_cont_private} is given by $ \{ x^\ast(x_o) \} = \{  (L+\Theta)^{-1} \Theta x_o  \} $ for each $ x_o $. 
\end{corollary}

\begin{example}
\label{ex:FJ}
An example of $ n=100 $ agents is shown in Fig.~\ref{fig:FJ_cont_private}. The introduction of $ h(\cdot) $ scrambles the initial conditions, as expected, see panel (c) of Fig.~\ref{fig:FJ_cont_private}. A level of privacy $ \lambda = 1$ is requested. Both $ x(t) $ and $ y(t) $ converge to the same $ x^\ast = (L+\Theta)^{-1} \Theta x_o $, see panel (d) of Fig.~\ref{fig:FJ_cont_private}, although neither now respects the rankings during the transient (i.e., unlike for \eqref{eq:FJ_cont}, for \eqref{eq:FJ_cont_private} it is no longer true that $ x_i(t_1) < x_j(t_1) $ $ \Longrightarrow $ $ x_i(t_2) < x_j(t_2) $ for all $ t_2> t_1$). 
 
\begin{figure*}[htb]
\begin{center}
\subfigure[]{
\includegraphics[angle=0, trim=1cm 7.5cm 9cm 1cm, clip=true, width=4.2cm]{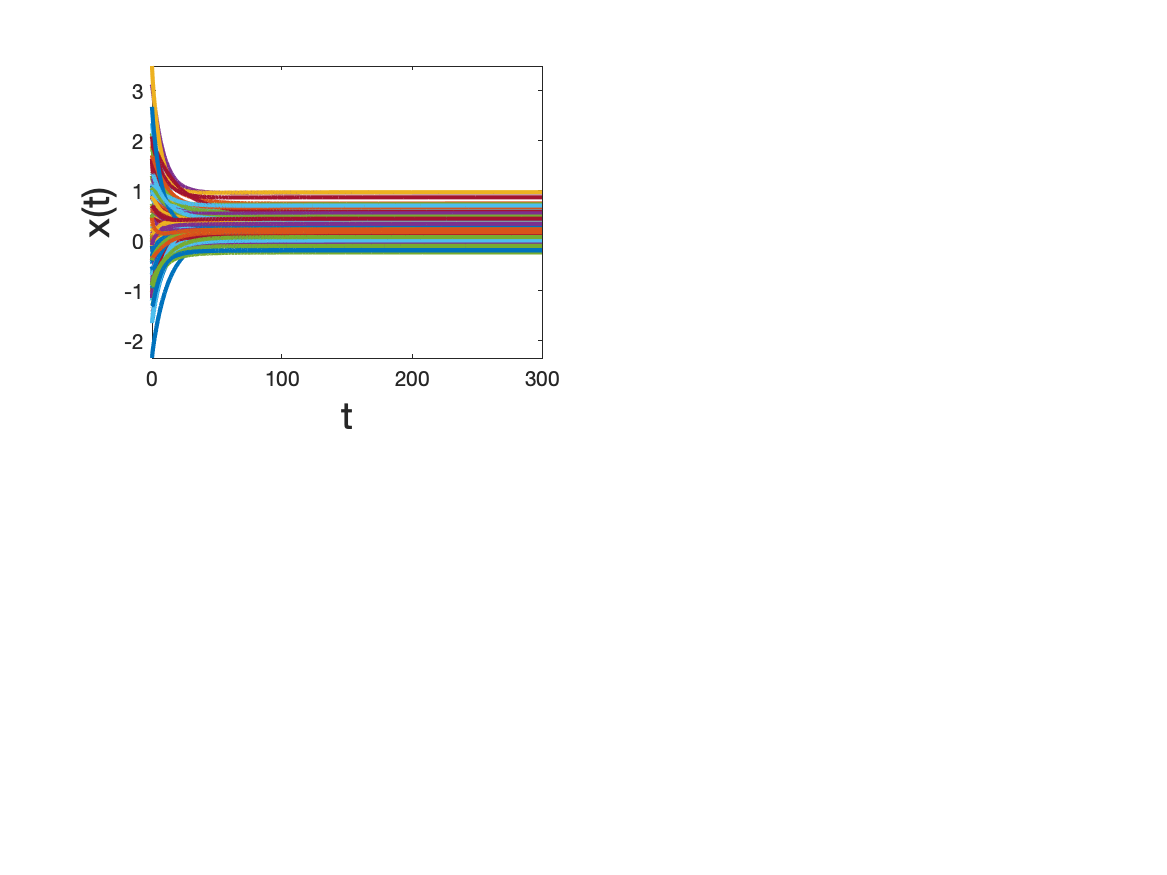}} 
\subfigure[]{
\includegraphics[angle=0, trim=1cm 7.5cm 9cm 1cm, clip=true, width=4.2cm]{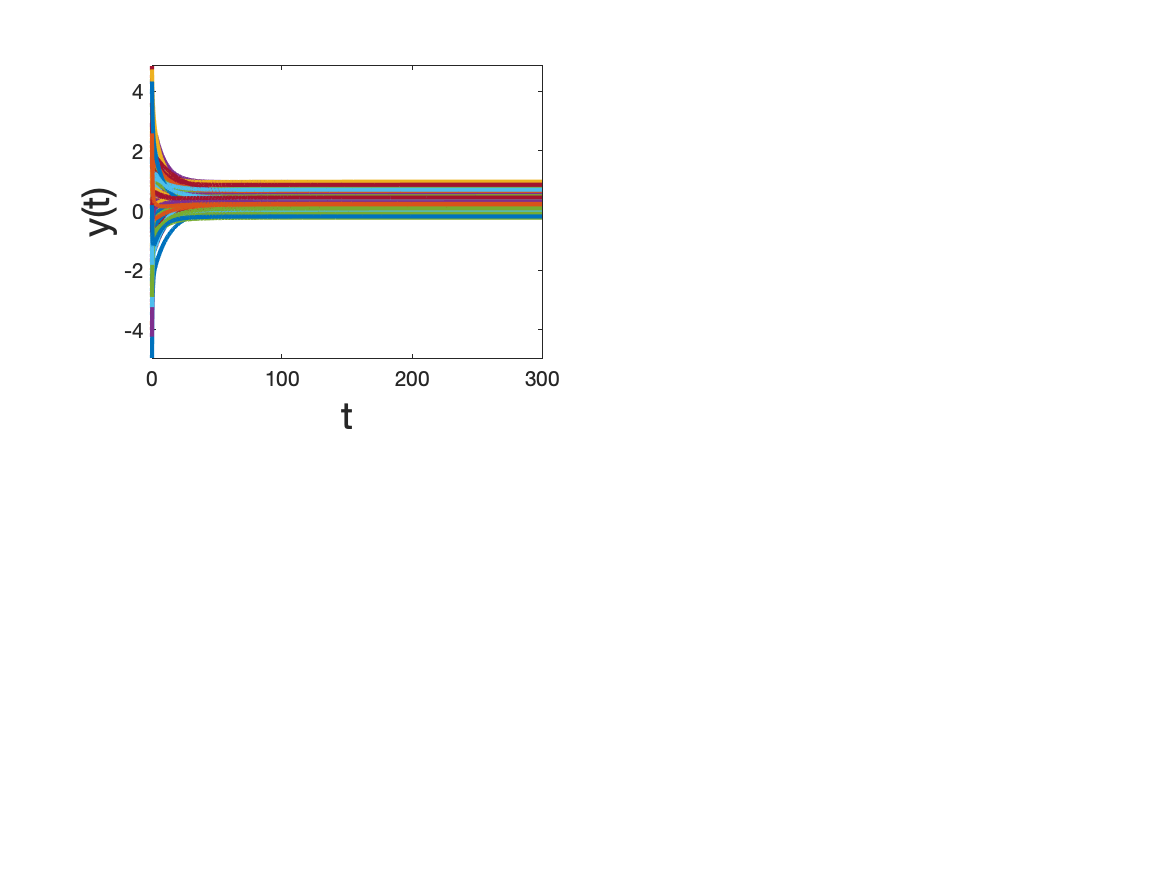}}
\subfigure[]{
\includegraphics[angle=0, trim=1cm 7.5cm 9cm 1cm, clip=true, width=4.2cm]{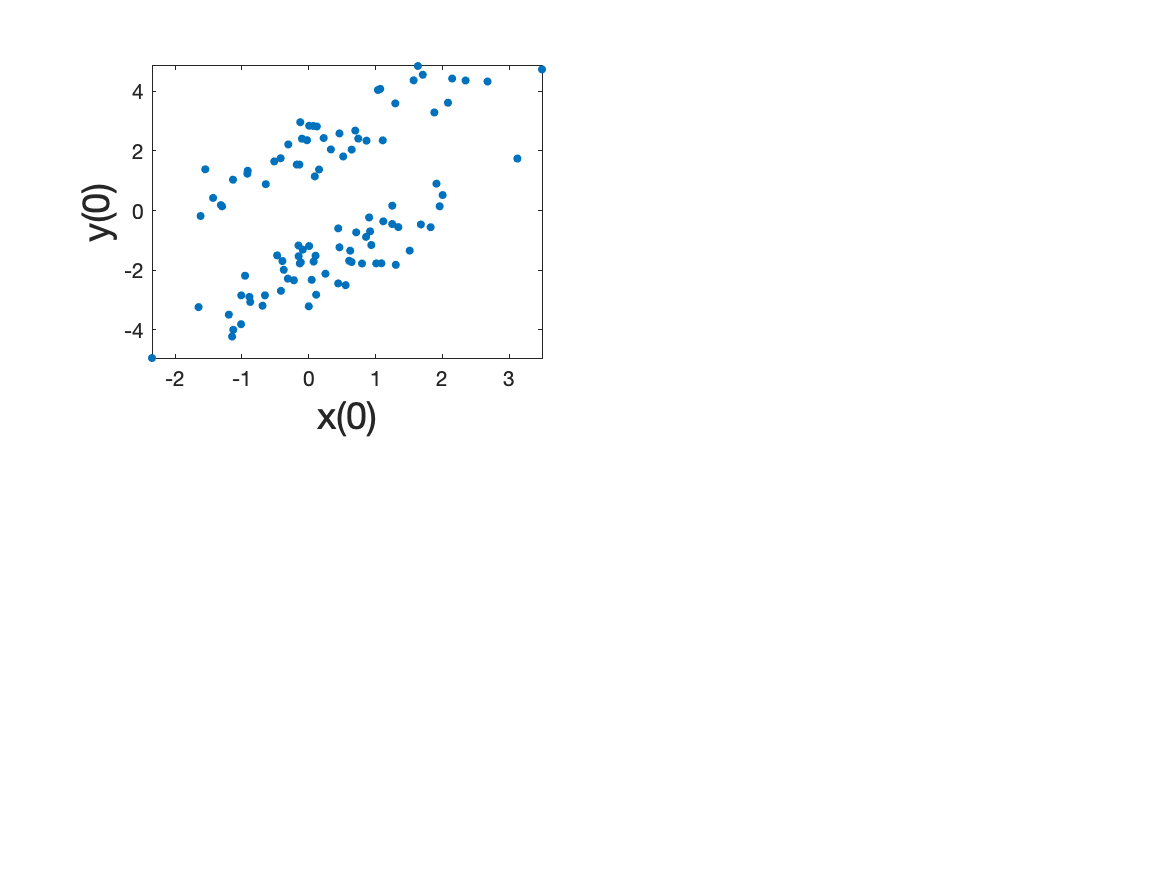}}
\subfigure[]{
\includegraphics[angle=0, trim=1cm 7.5cm 9cm 1cm, clip=true, width=4.2cm]{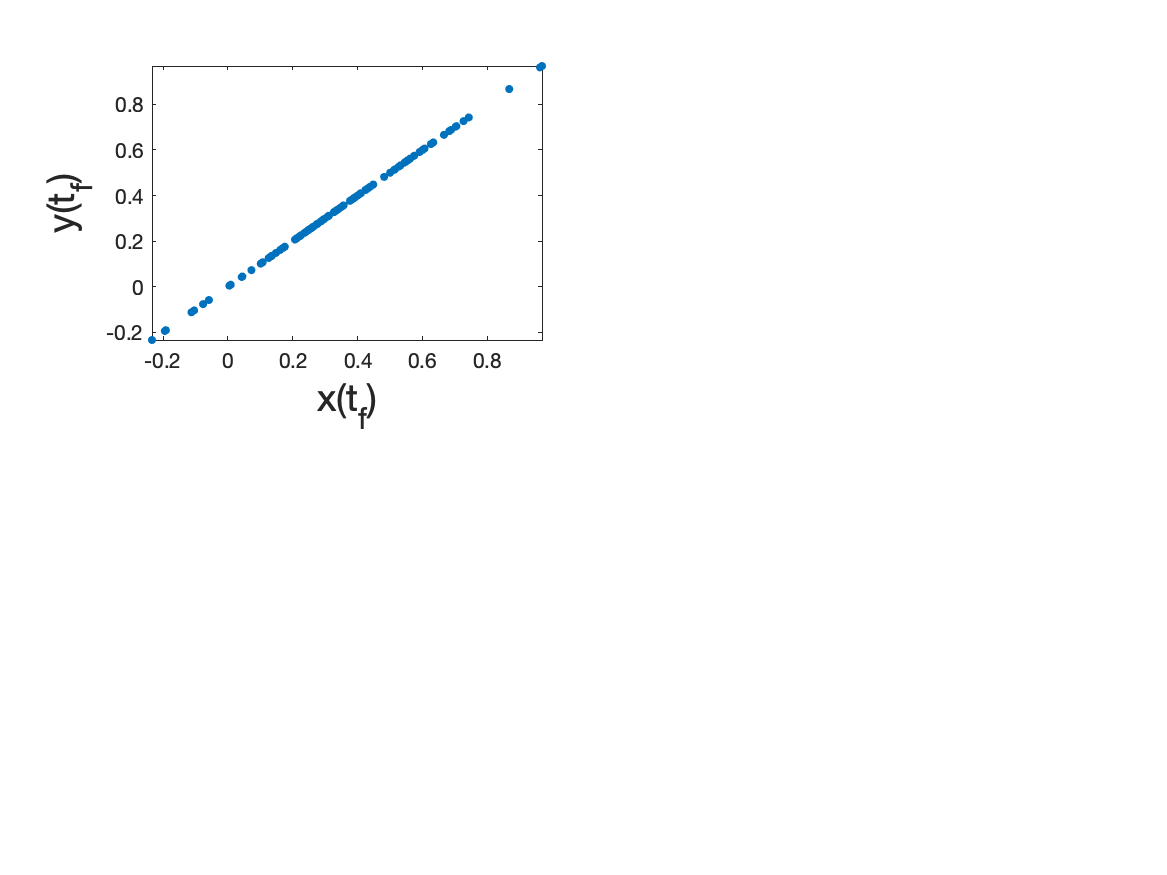}}
\caption{Privacy-preserving continuous-time Friedkin-Johnsen model of Example~\ref{ex:FJ}. (a): private state $ x(t)$; (b): masked output $ y(t)$; (c): initial condition $ x(0) $ vs. $ y(0)$; (d): final condition $ x(t_f) $ vs. $ y(t_f)$, where $ t_f = $ final time of the simulation.
}
\label{fig:FJ_cont_private}
\end{center}
\end{figure*}
\end{example}

%%%%%%%%%%%%%%%%%%%%%%%%%%%%%%%%%%%%%%%%%%

\section{Dynamically private average consensus}
\label{sec:av-consensus}

In the average consensus problem, $ f(x) = -L x $, with $ L$ a weight-balanced Laplacian matrix: $ L \bfone = L^T \bfone =0 $, with $ \bfone =\begin{bmatrix} 1 & \ldots & 1 \end{bmatrix}^T \in \mathbb{R}^n$.
When $L$ is irreducible, the equilibrium point is $ x^\ast(x_o) = (\bfone^T x_o/n) \bfone $. The system has a continuum of equilibria, described by $ {\rm span}(\bfone) $, and each $ x^\ast (x_o) $ is globally asymptotically stable in $ {\rm span} (\bfone)^\perp$, see \cite{Olfati2003Consensus}.

\begin{theorem}
\label{thm:av-consensus}
Consider the system
\beq
\dot x = - L x , \qquad x(0) =x_o 
\label{eq:consensus1}
\eeq
where $ L $ is an irreducible, weight-balanced Laplacian matrix, and denote $ \eta = \bfone^T  x_o /n $ its average consensus value.
Then $ x^\ast = \eta \bfone $ is a global uniform attractor on $ {\rm span} (\bfone)^\perp$ for the masked system 
\beq
\begin{split}
\dot x & = - L y  \\
y & = h(t, \,  x, \, \pi ) = \left(I + \Phi e^{-\Sigma t } \right)  \left( x + e^{-\Delta t} \gamma \right) .
\end{split}
\label{eq:consensus_xy}
\eeq
Furthermore, if Assumption~\ref{ass1} holds, then \eqref{eq:consensus_xy} is a dynamically private version of \eqref{eq:consensus1}.
\end{theorem}

Also in this case our masked system is an asymptotically autonomous time-varying system.
\begin{corollary}
\label{cor:consensus-autonomous}
The masked system \eqref{eq:consensus_xy} is asymptotically autonomous with \eqref{eq:consensus1} as limit system. 
The $ \omega$-limit set of \eqref{eq:consensus_xy} is given by $ \{ \eta  \bfone    \} $ for each $ x_o $. 
\end{corollary}

\begin{remark}
Even if  \eqref{eq:consensus1} has $ x^\ast =\eta \bfone  $ as a globally asymptotically stable equilibrium point in $ {\rm span}(\bfone)^\perp$, the masked system  \eqref{eq:model_xy} does not have equilibria because of the extra inhomogeneous term in the right hand side%of \eqref{eq:consensus2}
, hence we cannot talk about stability of $ \eta \bfone $. 
Nevertheless, $ x^\ast =\eta \bfone $ remains a global attractor for all trajectories of the system  in $ {\rm span}(\bfone)^\perp$.
\end{remark}

\begin{remark}
Since the evolution of the masked system \eqref{eq:consensus1} is restricted to the $ n-1$ dimensional subspace $ {\rm span}(\bfone)^\perp$, our masked consensus problem (as any exact privacy preserving consensus scheme) makes sense only when $ n>2$. The case $ n=2 $ never satisfies Assumption~\ref{ass1} when $L$ is irreducible.
\end{remark}

\begin{example}
\label{ex:private-consensus}
In Fig.~\ref{fig:consensus1} a private consensus problem is run among $ n=100 $ agents. 
Both $ x(t) $ and $ y(t) $ converge to the same consensus value $ \eta = \bfone^T x(0)/n$, but the initial condition $ y(0) $ does not reflect $ x(0)$, not even when $ x_i(0) $ is already near $ \eta $ ($h(\cdot ) $ does not preserve neighborhoods, see panel (c) of Fig.~\ref{fig:consensus1}). 
The level of privacy measure imposed in this simulation is $ \lambda=1 $. 
Notice that $ \bfone^T x(t)/n $ is constant over $t$, while $ \bfone^T y(t)/n $ is not, i.e., the output mask hides also the conservation law.
Notice further that a standard Lyapunov function used for consensus, like $ V_{mm}(t) = \max_i(x_i(t)) - \min_i (x_i(t)) $, does not work in our privacy-preserving scheme (see panel (d) of Fig.~\ref{fig:consensus1}), which reflects the fact that the system \eqref{eq:consensus_xy} is not asymptotically stable in $ {\rm span}(\bfone)^\perp $.
The convergence speed of the time-dependent part can be manipulated by selecting the factors $ \sigma_i $ and $ \delta_i $ appropriately.

\begin{figure*}[htb]
\begin{center}
\subfigure[]{
\includegraphics[angle=0, trim=1cm 7.5cm 9cm 1cm, clip=true, width=4.2cm]{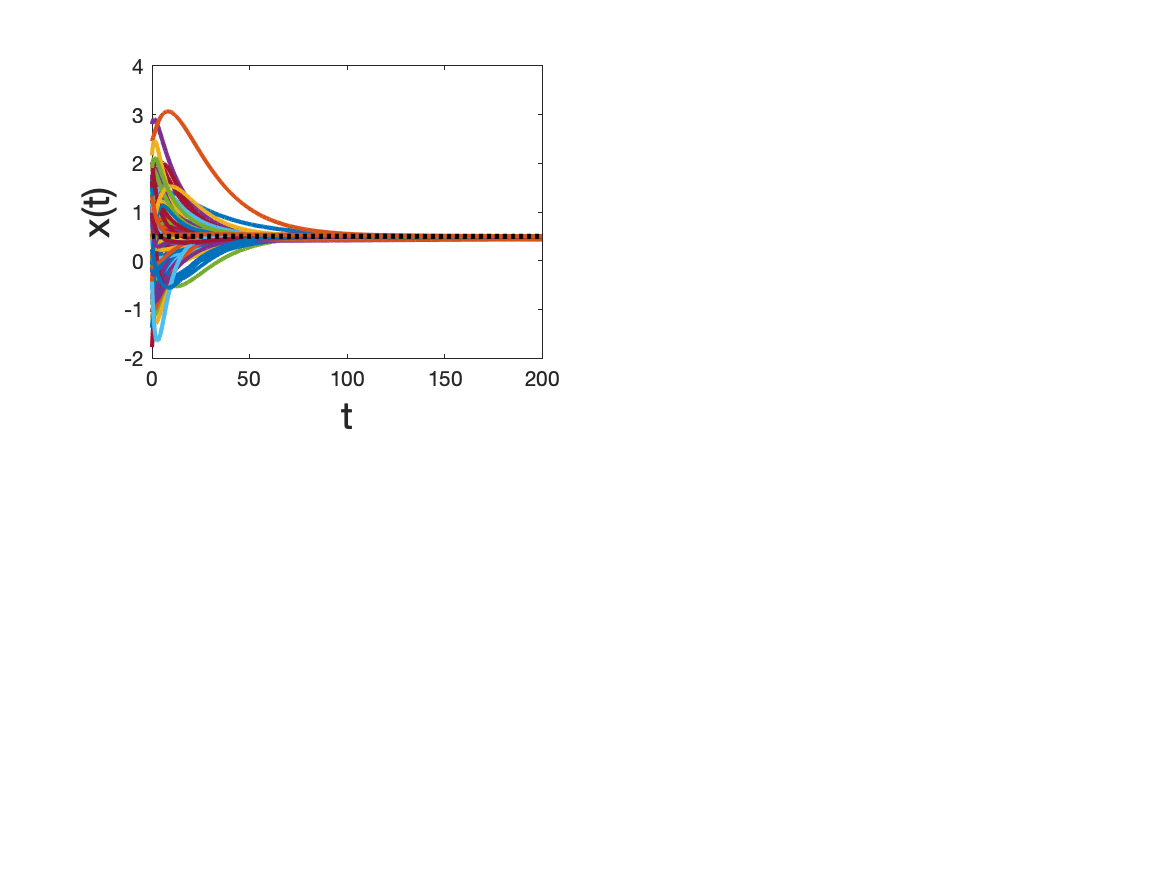}} 
\subfigure[]{
\includegraphics[angle=0, trim=1cm 7.5cm 9cm 1cm, clip=true, width=4.2cm]{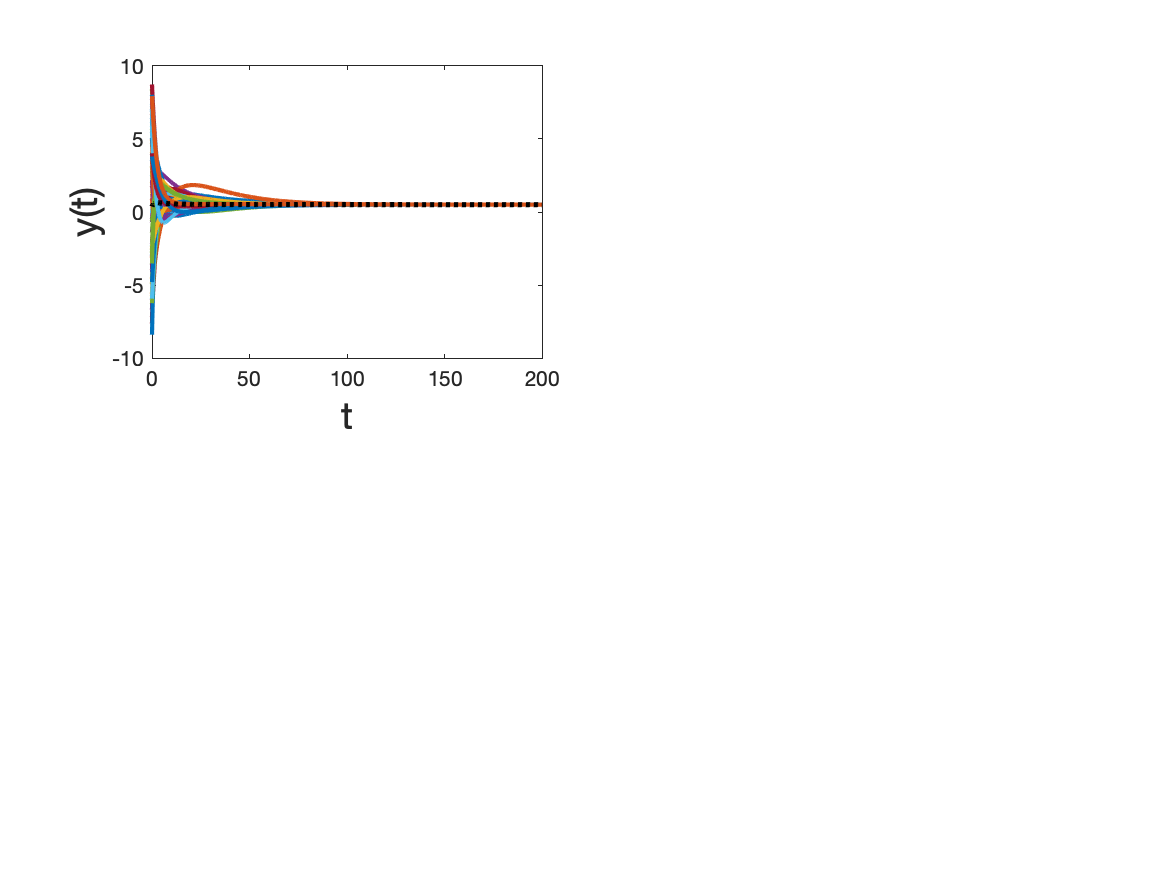}}
\subfigure[]{
\includegraphics[angle=0, trim=1cm 7.5cm 9cm 1cm, clip=true, width=4.2cm]{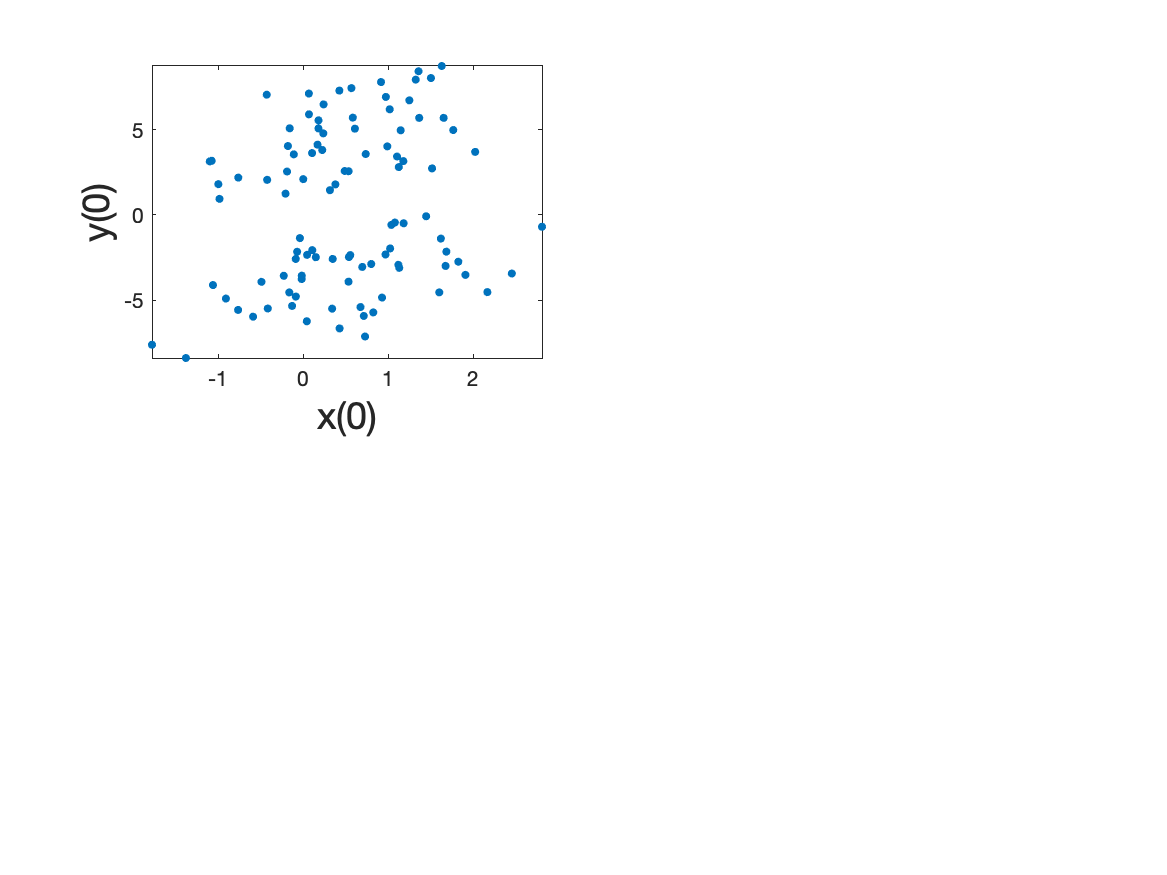}}
\subfigure[]{
\includegraphics[angle=0, trim=1cm 7.5cm 9cm 1cm, clip=true, width=4.2cm]{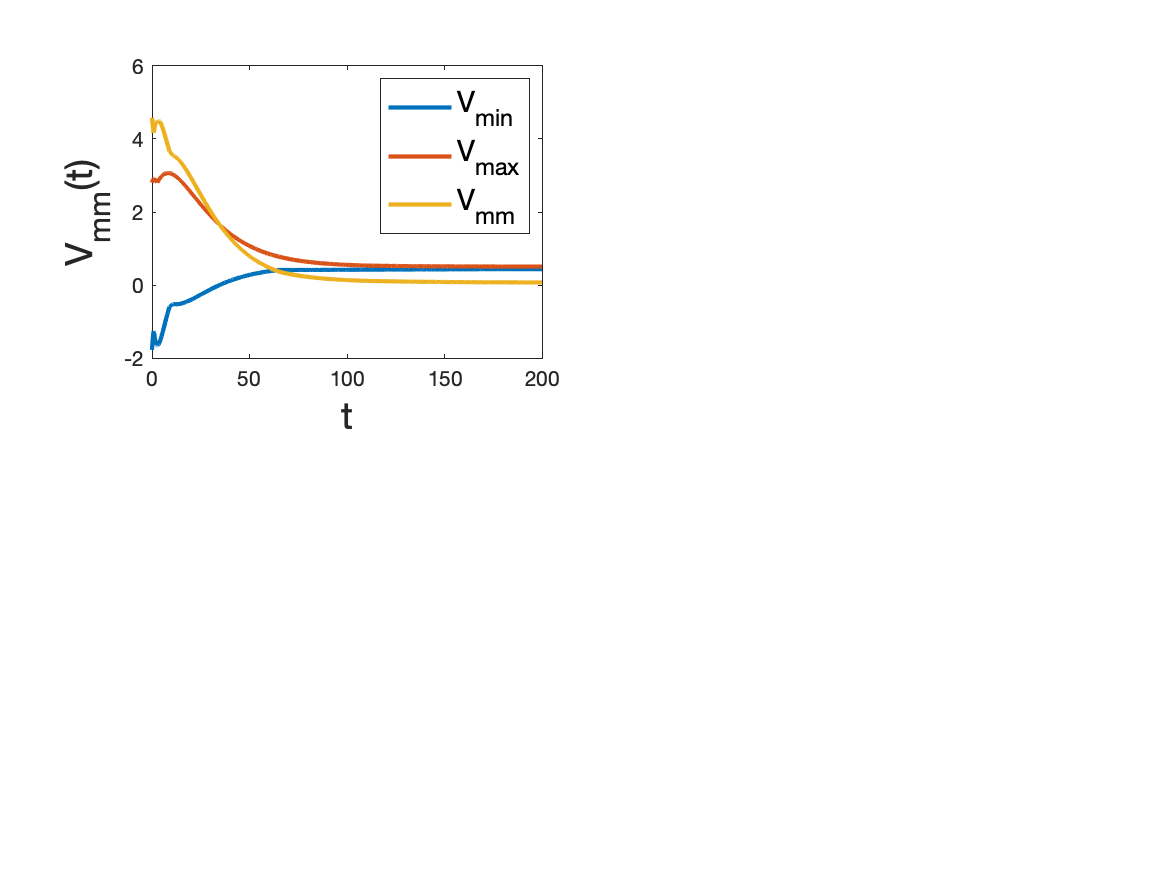}}
\caption{Privacy-preserving consensus of Example~\ref{ex:private-consensus}. (a): private state $ x(t)$; (b): masked output $ y(t)$; (c): initial conditions $ x(0) $ vs. $ y(0)$; (d): $ V_{mm}(t) = \max_i(x_i(t)) - \min_i (x_i(t)) $. The black dotted line in (a) resp. (b) represent $ \bfone^T x(t)/n $, resp. $ \bfone^T y(t)/n $.
}
\label{fig:consensus1}
\end{center}
\end{figure*}

\end{example}

%%%%%%%%%%%%%%%%%%%%%%%%%%%%%%%%%%%%%%%%%%

\section{Privacy for higher order systems: the case of pinned synchronization}
\label{sec:synchro}

When instead of a scalar variable, at each node we have a vector of variables $ x_i \in \mathbb{R}^\nu$, $ \nu >1 $, then the definition of output mask can be straightforwardly extended by defining $ h_i(t, x_i, \pi_i ) $ as a $ \nu$-dimensional diagonal map.  
For instance for the vanishing affine output mask, in place of \eqref{eq:van-aff-mask-scalar} at each node we can use 
\[
h_i (t, x_i, \pi_i ) = (I + \Phi_i e^{-\Sigma_i t } ) ( x_i + e^{-\Delta_i t } \gamma_i )
\]
where $ \Phi_i = {\rm diag}( \phi_{i,1}, \ldots, \phi_{i,\nu} ) $, $ \Sigma_i= {\rm diag}( \sigma_{i,1}, \ldots, \sigma_{i,\nu} ) $, $ \Delta_i = {\rm diag}( \delta_{i,1}, \ldots, \delta_{i,\nu }) $, and $ \gamma_i = \begin{bmatrix} \gamma_{i,1} & \ldots &  \gamma_{i,\nu} \end{bmatrix}^T $.
The formalism introduced in the paper extends unaltered.

We will now investigate privacy protection in a standard example of coordination of multivariable multiagent systems: synchronization via pinning control of identical nonlinear systems with diffusive couplings \cite{Chen4232574,Yu-doi:10.1137/100781699,ZHOU2008996}. 
Other settings of multiagent coordination can be treated in an analogous way. 

Consider a network of $n$ agents obeying the following set of coupled differential equations
\begin{align}
\dot x_i & = f(x_i) - \sum_{j=1}^n \ell_{ij} R x_j - p_i R (x_i - s) , \qquad i=1, \ldots, k 
\label{eq:synchro1} \\
\dot x_i & = f(x_i) - \sum_{j=1}^n \ell_{ij} R x_j, \qquad i=k+1, \ldots, n
\label{eq:synchro2}
\end{align}
where $ x_i \in \mathbb{R}^\nu $, $ L = (\ell_{ij} ) $ is an irreducible Laplacian matrix, and $ R$ is a symmetric positive definite matrix of inner couplings.
The extra term in the first $ k$ equations expresses the coupling with a pinned node ($ p_i = $ pinning gain), acting as an exosystem for \eqref{eq:synchro1}-\eqref{eq:synchro2} and obeying to the law
\beq
\dot s  = f(s) .
\label{eq:synchro3}
\eeq
The system \eqref{eq:synchro3} can represent an equilibrium point, a periodic or a chaotic system \cite{Yu-doi:10.1137/100781699}. 
Synchronization of \eqref{eq:synchro1}-\eqref{eq:synchro2} to the exosystem \eqref{eq:synchro3} corresponds to 
\[
\lim_{t\to\infty} \| x_i(t) - s(t) \| =0 \quad \forall \, x_i(0) \in \mathbb{R}^\nu, \quad  \forall \, i =1, \ldots, n.
\]
We need the following (standard) assumption:
\begin{assumption}
\label{ass2} {\rm (Global Lipschitzianity of the drift)}
$ f \,:\, \mathbb{R} \to \mathbb{R} $ is such that 
%\beq
%(x-z)^T (f(x) - f(z) ) \leq (x-z)^T Q R (x-z) \quad \forall \, x, \, z \in \mathbb{R}^\nu,
%\label{eq:synchro_ass1}
%\eeq
\beq
\| f(x) - f(z) \| \leq q \big( ( x-z)^T R (x-z) \big)^{\frac{1}{2}} \qquad \forall \; x, \, z \in \mathbb{R}^\nu
\label{eq:glob-Lipsc-pinning}
\eeq
for some positive constant $q$.
\end{assumption}
Under Assumption~\ref{ass2}, then a sufficient condition for global synchronization of \eqref{eq:synchro1}-\eqref{eq:synchro2} to \eqref{eq:synchro3} is given by the following matrix inequality
\beq
q \Xi \otimes R - \left( \frac{1}{2}(\Xi L + L^T \Xi ) + \Xi P \right) \otimes R  <0 
\label{eq:synchro_ass2}
\eeq
where $ \Xi = {\rm diag}(\xi ) $, with $ \xi = ( \xi_1, \ldots, \xi_n) $ the left eigenvector of $L$ relative to $0$, and $ P = {\rm diag}(p_1, \ldots, p_k, 0, \ldots, 0 ) $, see \cite{Yu-doi:10.1137/100781699} for more details.

\begin{theorem}
\label{thm:pinning}
Under the Assumptions~\ref{ass1} and~\ref{ass2}, if the solution $ s(t)$ of \eqref{eq:synchro3} is bounded $ \forall \, t \in [0, \, \infty) $, $ L $ is irreducible, and $ P $ is such that \eqref{eq:synchro_ass2} holds, then the exosystem \eqref{eq:synchro3} is a global attractor for the trajectories of the dynamically private system:
\begin{align}
\dot x_i & = f(y_i) - \sum_{j=1}^n \ell_{ij} R y_j - p_i R (y_i -s) , \quad i=1, \ldots, k 
\label{eq:synchro_masked1} \\
\dot x_i & = f(y_i) - \sum_{j=1}^n \ell_{ij} R y_j, \quad i=k+1, \ldots, n
\label{eq:synchro_masked2} \\
y_i & =  \left(I + \Phi_i e^{-\Sigma_i t } \right)  \left( x_i + e^{-\Delta_i t} \gamma_i \right) ,  \quad i=1, \ldots, n.
\label{eq:synchro_masked3}
\end{align}
\end{theorem}

\begin{remark}
Notice that the masked system \eqref{eq:synchro_masked1}-\eqref{eq:synchro_masked3} is not  asymptotically autonomous, as its limit system \eqref{eq:synchro1}-\eqref{eq:synchro2} is a function of the exosystem $ s(t)$ which also constitutes the $ \omega$-limit set of the system.
\end{remark}

\begin{example}
\label{ex:synchro}
Consider the case of an $ f(\cdot) $ representing a three dimensional chaotic attractor (here the model presented in \cite{ZHOU2008996} is used). 
In Fig.~\ref{fig:synchro1} a system of $ n=50$ coupled agents synchronize to an exosystem $ s(t) $ obeying the same law. The privacy measure in this example is set to $ \lambda=10$. The convergence speed can be tuned by changing the $ \Sigma_i$ and $ \Delta_i $ parameters of the masks.
\begin{figure*}[htb]
\begin{center}
\subfigure[]{
\includegraphics[angle=0, trim=1cm 7.5cm 9cm 1cm, clip=true, width=4.2cm]{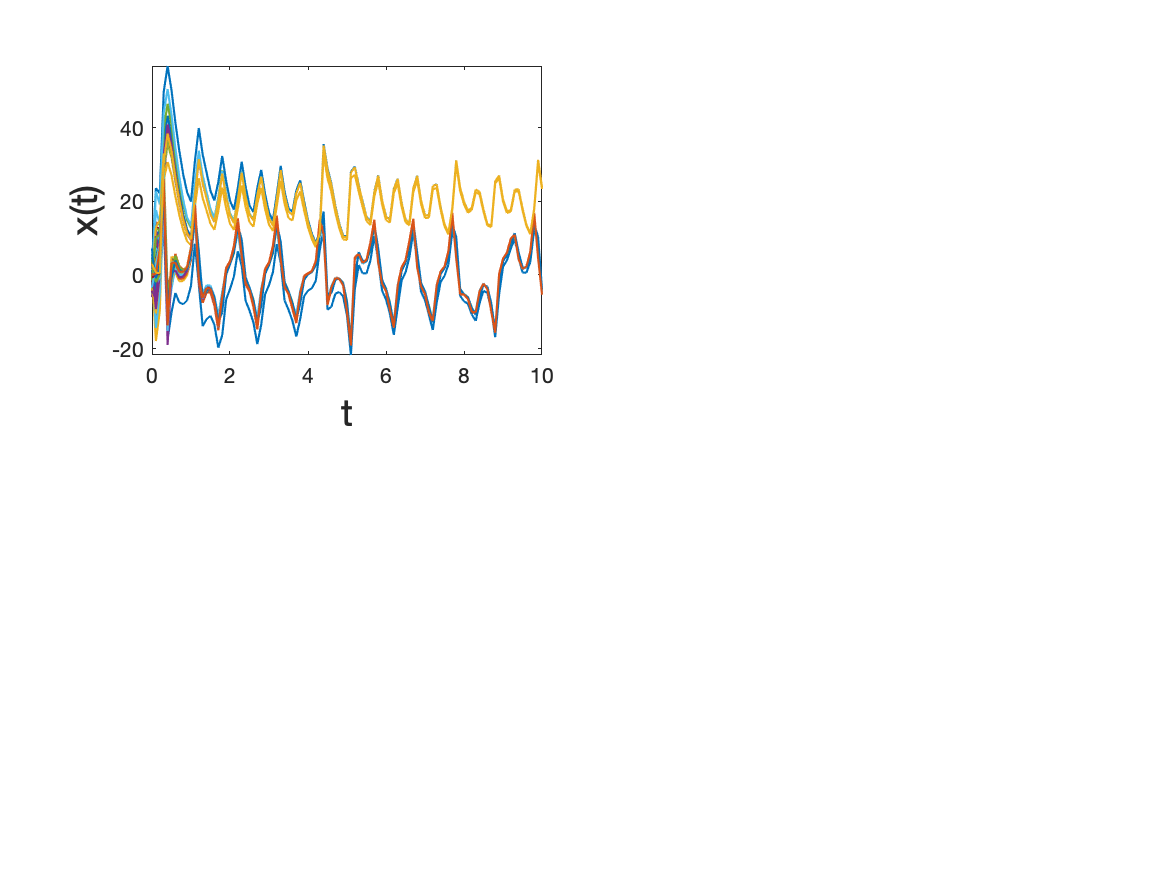}} 
\subfigure[]{
\includegraphics[angle=0, trim=1cm 7.5cm 9cm 1cm, clip=true, width=4.2cm]{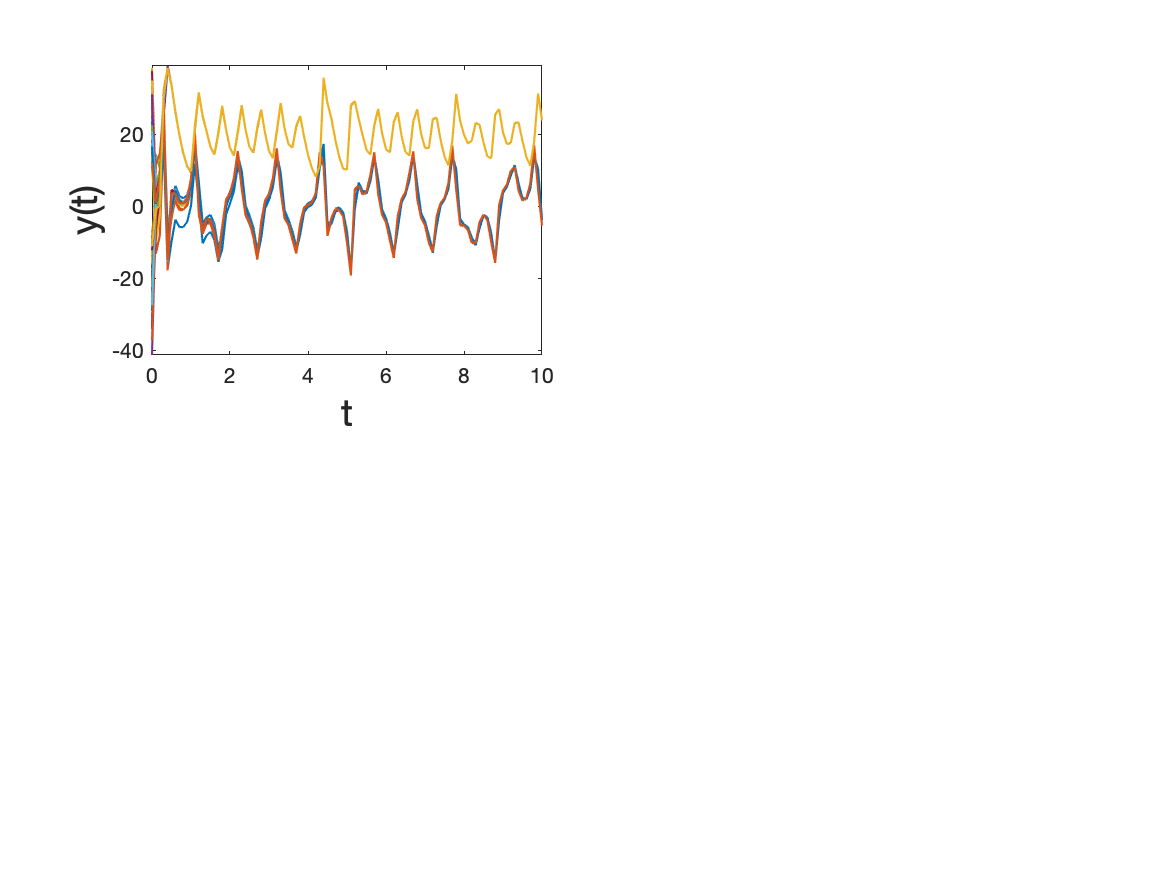}}
\subfigure[]{
\includegraphics[angle=0, trim=1cm 7.5cm 9cm 1cm, clip=true, width=4.2cm]{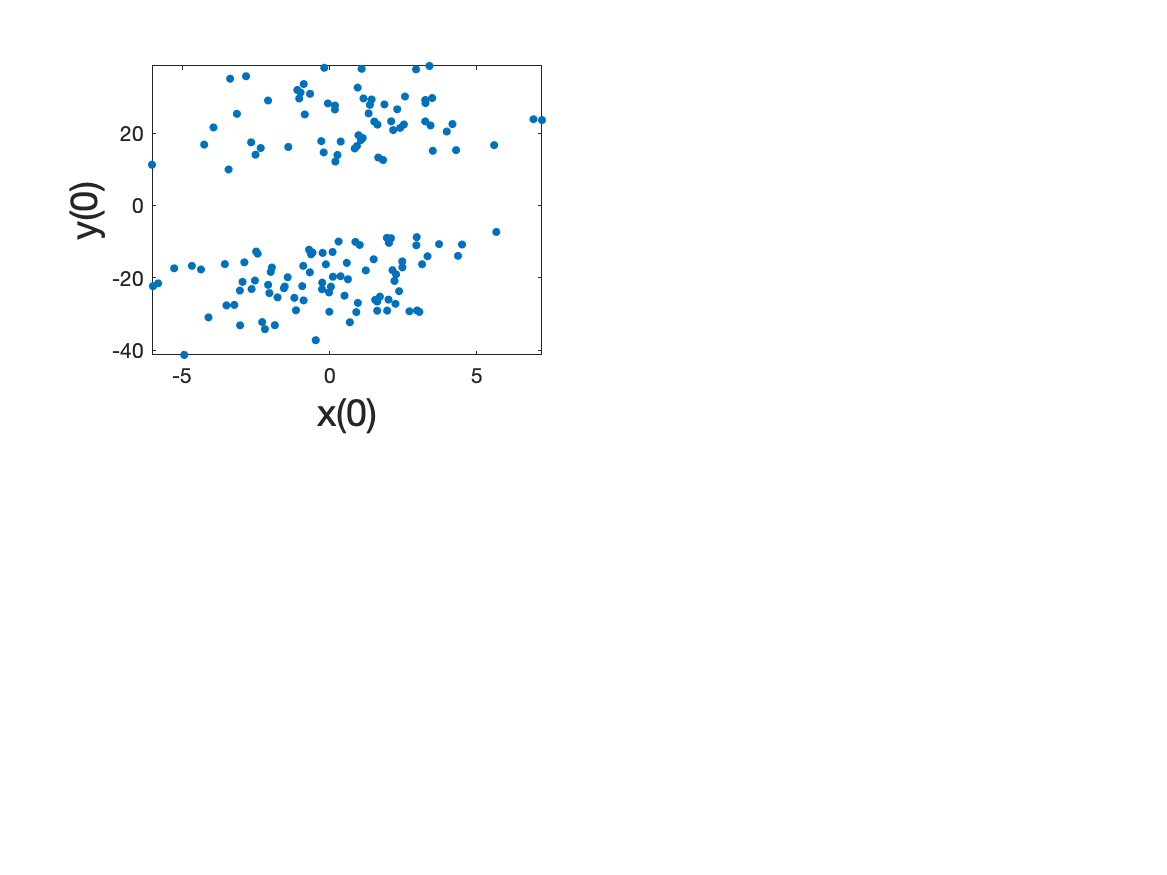}}
\subfigure[]{
\includegraphics[angle=0, trim=1cm 7.5cm 9cm 1cm, clip=true, width=4.2cm]{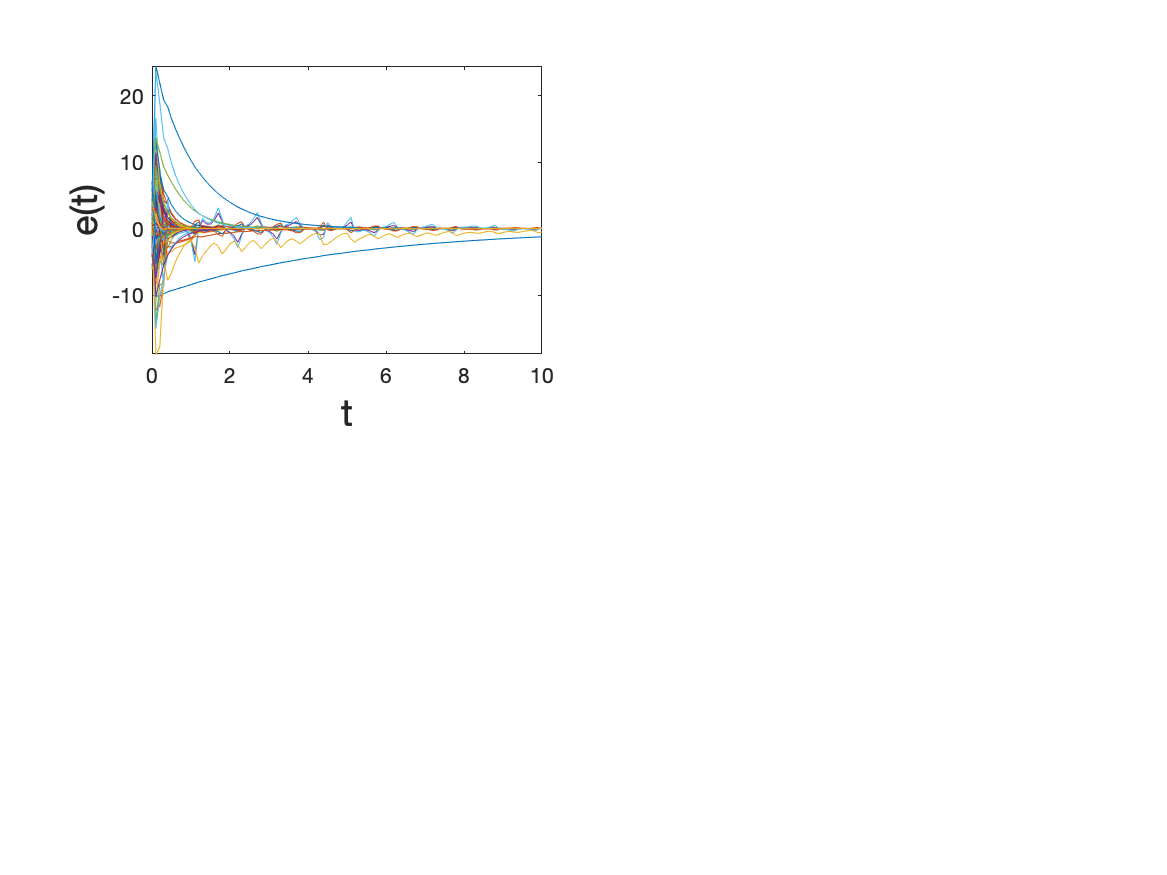}}

\caption{Privacy-preserving pinned synchronization of Example~\ref{ex:synchro}. (a): private state $ x(t)$; (b): masked output $ y(t)$; (c): initial condition $ x(0) $ vs. $ y(0)$; (d): error $ e(t)$.
}
\label{fig:synchro1}
\end{center}
\end{figure*}

\end{example}

\section{Conclusions}

The approach to privacy protection we have taken in this paper is exact and inspired by classical nonlinear systems techniques. 
While most of the assumptions under which it holds are fairly simple and reasonable (only the internal state of an agent and the parameters of its output mask must be kept private), the need to have non completely covering neighborhoods (Assumption~\ref{ass1}) is instead restrictive, but difficult to dispense with without requiring some other form of restriction (for instance privacy of the vector fields themselves). 
Assumption~\ref{ass1} is key to guarantee the impossibility for an eavesdropper to identify a model of the system, and hence to set up an observer for $ x_o$.
Notice that a breaching of the privacy at one node does not compromise the other nodes. 

From a system-theoretical perspective, the most interesting fact described in the paper is that privacy seems incompatible with a point being a fixed point of a dynamical system,  as in that case if all agents happen to have initial conditions already on the fixed point, privacy is compromised (an agent will see the same stationary messages being exchanged among its neighboring nodes for all $ t$). 
By extension of the same argument, approximate privacy (at any level of accuracy) does not seem to be compatible with stability.
It is intriguing to investigate if concepts like $ \epsilon$-differential privacy \cite{Cortes7798915} can be rephrased in these more dynamical terms.

Several generalizations of our approach are possible.
First of all an equivalent framework for discrete-time systems should be developed.
Then it is easy to think of output masks that vanish in finite time rather than asymptotically. 
More complicated seems to be integrating the time dependence introduced by an output mask with a time-varying communication graph. 
Even more challenging is the case in which, instead of global exponential stability (perhaps on ``slices'' of the state space if there is a continuum of equilibria) of the unmasked system, this last has multiple isolated locally exponentially stable equilibria. 
In this case even a transient output mask may lead to tipping over from one basin of attraction to another, hence it should be used with care.

\section{Acknowledgments}
The author would like to thank Claudio De Persis for useful discussions on the topic of the paper and the anonymous reviewers for constructive criticisms. This paper is dedicated to the memory of the author's father.

\appendix

%---------------------------------------------------------------------------
\section{Appendix}

\subsection{Auxiliary Lemmas}

The following lemma is inspired by \cite{MuASJC}, Thm 2.1 and \cite{saberi1990global}, Prop.~5, and provides us with a suitable comparison function to be used later in the paper.
% The proof of this Lemma and of all other results is in the Appendix.

\begin{lemma}
\label{lemma:comparison1}
Consider the scalar system 
\beq
\dot v = -\alpha (v) + \beta (v, t) + \zeta (t) , \qquad  v(t_o) = v_o \geq 0.
\label{eq:comparison1}
\eeq
If $ \alpha (v) \in \mathcal{K}_\infty^2 $, $  \beta \in \mathcal{KL}_\infty^{1,e} $ and $ \zeta \in \mathcal{L}^e$, then the solutions of \eqref{eq:comparison1} are all prolongable to $ \infty $  and bounded $ \forall \; v_o \geq 0$ and $\forall t_o \geq 0$. Furthermore,
\[
\lim_{t\to \infty } v(t) = 0 \qquad \forall \; v_o\geq 0, \quad \forall \; t_o \geq 0.
\]
\end{lemma}

\proof It follows from  $ \alpha(t) \geq 0$, $ \beta(v, t) \geq 0 $, $ \zeta(t) \geq 0$ for all $ v>0 $ and $ \alpha (0) =0 $ that $ \mathbb{R}_+ $ is invariant for the system \eqref{eq:comparison1}. 
If we can show that \eqref{eq:comparison1} remains bounded for all times, then \eqref{eq:comparison1} is also forward complete for all $ v_o\geq 0$.
Express $ \alpha(v) \in \mathcal{K}_\infty^2 $ as $ \alpha(v) = a v^2 $, $ \beta(v, t) \in \mathcal{KL}_\infty^{1,e} $ as $ \beta(v, t) = b v e^{-\delta_1 t}   $ and $ \zeta \in \mathcal{L}^e$ as $ \zeta(t) = c e^{-\delta_2 t} $ for some $ a, \, b, \, c >0$. 
Informally, boundedness follows from the fact that the globally exponentially stable ``unperturbed'' system $ \dot v = - a v^2 $ has a higher order as $ v\to \infty $ than the ``perturbation'' $ b v e^{-\delta_1 t } + c e^{-\delta_2 t }$. 
More in detail, in $ \mathbb{R}_+$, for $ v>1 $ it is $ v < v^2 $, hence we can write 
\[
b v e^{-\delta_1 t } + c e^{-\delta_2 t } < ( b  e^{-\delta_1 t } + c e^{-\delta_2 t }) v \qquad \forall \, v>1, \quad \forall \, t\geq t_o
\]
or 
\[
\dot v < ( -a v + b e^{-\delta_1 t } + c e^{-\delta_2 t }) v \qquad \forall \, v>1, \quad \forall \, t\geq t_o
\]
meaning that for $ v > \max \left( 1, \, \frac{b e^{-\delta_1 t_o } + c e^{-\delta_2 t_o } }{a} \right) $ it is $ \dot v <0$, $ \forall \, t \geq t_o $, i.e., the solution of \eqref{eq:comparison1} remains  bounded $ \forall \, v_o\geq 0 $ and $ \forall \, t_o \geq 0$.

Furthermore, $ \beta(v, t) $ and $\zeta(t) $ continuous, decreasing in $ t$ with $ \beta(v, t) {\to 0} $ and $ \zeta(t) \to 0 $ as $ t \to \infty$, imply also that for any $ v_o>0 $ there exists a $ t_1 \geq t_o $ such that $ \forall \; t > t_1 $ $ \dot v(t) <0$.
Together with $ \mathbb{R}_+$-invariance, this implies that $ \lim_{t\to\infty} v(t) = d \geq 0$. 
To show that it must be $ d=0 $, %we can use the same argument as in Lemma~2.3 of \cite{MuASJC}.  
let us assume by contradiction that $ d>0$. Then 
\[
\lim_{t\to \infty} \dot v(t) = \lim_{t\to \infty} (-\alpha (v) + \beta(v, t) + \zeta(t) ) = - \alpha(d) <0,
\]
meaning that there exists a $ t_2 > t_1 $ and a $ k \in (0, \, 1 ) $ such that 
\[
\dot v (t) < - k \alpha(d) <0 \qquad \forall \, t\geq t_2 .
\]
Applying the mean value theorem, we then have that $ \exists $ $ \tau \in [t_2, \, t] $ such that 
\[
\frac{v(t) - v(t_2)}{t-t_2} = \dot v (\tau) <  - k \alpha(d) <0 \qquad \forall \, t\geq t_2 ,
\]
from which it follows
\[
v(t) < - k \alpha(d) (t-t_2) + v(t_2) <0 \qquad \forall \, t\geq t_2 ,
\]
which is a contradiction since $ v\geq 0$.
\qed

%---------------------------------------------------------------------------
With Lemma~\ref{lemma:comparison1} in place, we can easily obtain the following sufficient condition for global convergence to the origin of a time-varying system in which the Lyapunov function has time derivative that is sign indefinite but bounded above by $ \mathcal{KL}_\infty^{1,e}  $ and $ \mathcal{L}^e$ functions, i.e., by terms growing linearly in the norm of the state and decaying exponentially in time. 
\begin{lemma}
\label{lemma:comparison2}
Assume that in the time-varying system \eqref{eq:ode_f(t,x)} $ g\, : \, \mathbb{R}_+ \times \mathbb{R}^n \to \mathbb{R}^n $ is such that the solution of \eqref{eq:ode_f(t,x)} exists unique in $ [t_o, \infty ) $ $ \forall \; x_o \in \mathbb{R}^n$ and $ \forall \; t_o \geq 0$.
If there exists a continuously differentiable function $ V(t, \, x) \, : \, \mathbb{R}_+ \times \mathbb{R}^n \to \mathbb{R} $, three $ \alpha_1$, $ \alpha_2$, $\alpha_3 \in \mathcal{K}_\infty^2 $, $ \beta \in \mathcal{KL}_\infty^{1,e}  $ and $ \zeta \in \mathcal{L}^e $ such that 
\begin{align}
 \alpha_1(\| x\| ) & \leq V(t, x ) \leq \alpha_2 (\| x \| ) \label{eq:compar_1}   \\
 \pde{V}{t} + \pde{V}{x} g(t, x) & \leq - \alpha_3 (\| x \| ) + \beta(\| x\|, t-t_o) + \zeta (t - t_o ) \label{eq:compar_2} 
\end{align}
$\forall $ $ t\geq t_o $, $ t_o\geq 0 $ and $ x_o \in \mathbb{R}^n$, then any solution of \eqref{eq:ode_f(t,x)} converges to $0$ uniformly in $ t_o$ as $ t\to \infty$.
\end{lemma}

\proof The right-hand side of \eqref{eq:compar_2} has the same structure as that of \eqref{eq:comparison1}, meaning we can apply the comparison lemma, using \eqref{eq:comparison1} with initial condition $ v(t_o) = V(t_o, x_o ) $.
Denoting $ v(t) $ the corresponding solution, it is then 
\beq
V(t, x ) \leq v(t)\qquad  \forall \; t \geq t_o .
\label{eq::compar_4}
\eeq
From Lemma~\ref{lemma:comparison1} and \eqref{eq::compar_4}, it follows that for all $ x_o $ it must be $ \lim_{t\to\infty} V(t, x(t)) =0 $ for any $ t_o \geq 0 $, hence from \eqref{eq:compar_1} $ \lim_{t\to\infty} \alpha_1(x(t)) =0 $ or $ \lim_{t\to\infty} x(t) =0$.
\qed

\begin{remark}
The sufficient conditions of Lemma~\ref{lemma:comparison1} (and hence of Lemma~\ref{lemma:comparison2}) can be rendered more general using for instance the  properties of  input-to-state stability \cite{SONTAG1995351}, or of cascade nonlinear systems \cite{PANTELEY2001Growth}.
\end{remark}

% \subsection{Proof of   Lemma~\ref{lemma:bound-affine-mask}}

\begin{lemma}
\label{lemma:bound-affine-mask}
For the output mask \eqref{eq:output-mask-affine-II}, it holds:
%\begin{align}
%\| x \| - \zeta (t) & \leq  \| y\| \leq  k ( \| x \| + \zeta (t) ) \label{eq:norm-y-1} \\
\beq 
\frac{\| y \|}{k} - \zeta (t)  \leq \| x\|  \leq   \| y \| + \zeta (t)  \label{eq:norm-y-2}
\eeq
%\end{align}
where $ k = \| I + \Phi \| $ and $ \zeta(t) =\| e^{-\Delta t } \| \in \mathcal{L}^e$.
\end{lemma}

\proof The inverse of \eqref{eq:output-mask-affine-II} is
\begin{align}
x  =  (I + \Phi e^{-\Sigma t} )^{-1}  y  - e^{-\Delta t } \gamma 
\label{eq:inverse-mask-affine-II-1} 
%\\
% & = (I - \Phi e^{-\Sigma t } (I + \Phi e^{-\Sigma t} )^{-1})   y - e^{-\Delta t } \gamma .
% \label{eq:inverse-mask-affine-II-2}
\end{align}
Notice that $ \| I + \Phi e^{-\Sigma t} \| \leq \| I + \Phi  \| =k$. 
We have from \eqref{eq:output-mask-affine-II} and from the definition of $ \zeta(t)$ that 
\[
\| y \| \leq k ( \| x \| + \zeta(t) ) ,
 \] 
and from \eqref{eq:inverse-mask-affine-II-1}, 
\[
\| x \| \leq \|y\| + \zeta(t) .
\] 
The bounds \eqref{eq:norm-y-2} follow combining these two inequalities. 
\qed

%---------------------------------------------------------------------------

\subsection{Proof of   Proposition~\ref{prop:metric}}
For \eqref{eq:ex-additive-mask} it is $ \rho_i (x_{o,i}) = | \gamma_i | $, hence it is enough that each agent chooses the parameter $ \gamma_i $ such that $ |\gamma_i | >\lambda $, independently of $ x_{o,i} $. 
For  \eqref{eq:ex-affine-mask}, $ \rho_i (x_{o,i} ) = |(c_i - 1) x_{o,i} + \gamma_i | >\lambda $ is satisfied for an infinite number of parameter pairs $ (c_i, \, \gamma_i ) $ with $ c_i >1 $ and $ \gamma_i \neq 0 $. When this inequality is satisfied for all agents then $ \rho(x_o) >\lambda$.
Similarly, for \eqref{eq:van-aff-mask-scalar} there exist infinitely many parameter pairs $ (\phi_i, \, \gamma_i)$ with $ \phi_i >0 $ and $ \gamma_i \neq 0 $ satisfying $ \rho_i  (x_{o,i}) = |\phi_i x_{o,i} + (1+ \phi_i ) \gamma_i | >\lambda$. 
\qed

%---------------------------------------------------------------------------

\subsection{Proof of   Proposition~\ref{prop:ass1:indiscern}}

\proof
Consider an eavesdropping agent $ j$ trying to discern the initial condition of agent $i$, i.e., trying to estimate $ x_{o,i}$ based on K1 and K2. 
This requires to compute the state of $i$ from the available outputs.
From \eqref{eq:model_xyi}, there are two possible ways to proceed. 
The first is to proceed ``statically'' by inverting $ h_i (\cdot ) $ in \eqref{eq:model_xyi_b}, and the second to proceed dynamically using both \eqref{eq:model_xyi_a} and \eqref{eq:model_xyi_b}.
Concerning the first possibility, from \eqref{eq:model_xyi_b} computing $ x_i(t, x_{o,i} ) $ from $ y_i (t, x_{o,i}) $ requires to invert the masked map $ h_i (\cdot) $ for each $t$. 
From U1, however, this inversion is not possible for agent $ j$, because $ h_i (\cdot ) $ is a privacy mask unknown to agent $j$. 
Concerning the second possibility, there are two possible options: the first is to use \eqref{eq:model_xyi_a} and \eqref{eq:model_xyi_b} to set up a system identification problem for $ h_i (\cdot ) $ and $ \pi_i $. 
However, from Assumption~\ref{ass1}, only a proper subset $ \mathcal{M}_{ij} =\{ \mathcal{N}_i\cup \{ i \} \} \cap \{ \mathcal{N}_j \cup \{ j \}\} $ of all output trajectories entering into $ f_i (\cdot ) $ (i.e., $\{ \mathcal{N}_i\cup \{ i \} \}$) is available to agent $j$.
Combining this with the essential neighborhood assumption \eqref{eq:non-trivial-neigh}, we obtain that agent $j$ cannot correctly compute the right hand side of \eqref{eq:model_xyi_a}, hence a system identification problem for \eqref{eq:model_xyi}  cannot be solved correctly. 
The second dynamical option is instead to consider the formal solution of \eqref{eq:model_xyi_a}
\[
x_i (t) = x_{o,i} + \int_0^t f_i(y_i, \, y_k , \, k \in \mathcal{N}_i ) d\tau .
\label{eq:discern-t}
\]
Since $ y^\ast = \lim_{t\to\infty} y(t) = \lim_{t\to\infty} x(t) = x^\ast$, if $ i \in \mathcal{N}_j$, when $ t\to\infty $,  $ x_i^\ast $ is available to $j$, hence $ x_{o,i} $ can be expressed as
\beq
 x_{o,i} = x_i^\ast-  \int_0^\infty f_i(y_i, \, y_k , \, k \in \mathcal{N}_i ) d\tau .
\label{eq:discern-infty}
\eeq
Also in this case, however, Assumption~\ref{ass1} combined with the essential neighborhood assumption \eqref{eq:non-trivial-neigh} implies that agent $ j$ cannot correctly estimate the integral in \eqref{eq:discern-infty}, as $ \mathcal{M}_{ij}  \nsubseteq \{ \mathcal{N}_i \cup \{ i \}\} $.
In summary, since neither static nor dynamical methods for estimating $ x_{o,i}$ can be applied, we can conclude that the initial condition $ x_{o,i}$ is indiscernible for agent $ j$. Since Assumption~\ref{ass1} is valid for all agents, we can also conclude that $ x_o $ is indiscernible, and therefore that \eqref{eq:model_xy} is a dynamically private version of \eqref{eq:model}.
\qed

% -----------------------------------------

\subsection{Proof of   Proposition~\ref{prop:no-equil}}
The right hand side of the dynamics in \eqref{eq:model_xy} is autonomous.
Assume there exists $ y^\ast $ such that $ f(y^\ast) =0$. 
Since, from P4 of Definition~\ref{def:privacy_mask}, $ h(\cdot ) $ is invertible in $ x$ for each $ t$, by the implicit function theorem, there exists an $ x^\ast (t) $ such that $ y^\ast = h(t, x^\ast(t), \pi) $. 
If $ x^\ast(t) $ is time-varying, then it is not an equilibrium point for \eqref{eq:model_xy}. 
If instead $ x^\ast $ is time-invariant, then, from $ \lim_{t\to \infty} y(t) = \lim_{t\to \infty} x(t) $, it must be $ x^\ast = y^\ast$. 
But then, choosing $ x(0) = x^\ast$, it is $ y^\ast = h(0, x^\ast, \pi) = x^\ast $, i.e., P2 of Definition~\ref{def:privacy_mask} is violated, hence also this case cannot happen in a privacy mask.
%\qed
%%---------------------------------------------------------------------------
%

%\subsection{Proof of   Proposition~\ref{prop:asymtp-autonomous}}
As for the second part, we need to show that $ f(h(t, x, \pi)) \to f(x) $ as $ t \to \infty $ uniformly on compacts of $ \mathbb{R}^n $ \cite{Artstein1976}.
From P5 and $ h \in C^1 $, there exists an increasing, diverging sequence $ \{ t_k \} $ for which $ h_i (t_k,  x_i, \pi_i ) \to x_i $ as $ t_k \to \infty $, i.e., pointwise convergence holds.
In particular, for any $ \epsilon >0$, from pointwise convergence, there exists a $ \nu_o (x_i) $ such that, for all $ \nu>\nu_o $, $ |  h_i (t_{\nu}, x_i, \pi_i) - x_i | <\epsilon/2 $.
Pick two indexes $ \nu_1 = \nu_1(x_i) $, $ \nu_2=\nu_2(x_i) $ such that $ \nu_m > \nu_o$, $m = 1, \, 2 $. 
Then $ |  h_i (t_{\nu_1}, x_i, \pi_i) -  h_i (t_{\nu_2}, x_i, \pi_i)  | \leq |  h_i (t_{\nu_1}, x_i, \pi_i) - x_i | +  |  h_i (t_{\nu_2}, x_i, \pi_i) - x_i | \leq\epsilon/2 + \epsilon/2$. 
Selecting $ \nu_s = \sup_{x_i\in \mathcal{X}_i } \left\{ \nu_m (x_i) , \; m=1, \, 2 \right\}  $, then the Cauchy condition for uniform convergence applies and we have for any  integer $ \mu $ 
\[
\begin{split}
&  |  h_i (t_{\nu_s}, x_i, \pi_i) - x_i | \\
& \quad = \lim_{\mu \to \infty}  |  h_i (t_{\nu_s}, x_i, \pi_i) -  h_i (t_{\nu_s+ \mu}, x_i, \pi_i)  | \leq \epsilon .
\end{split}
 \]
 Hence, for a certain subsequence $ \{ t_\nu\} $ of $ \{ t_k \}$ it is $ {\rm sup}_{x_i \in \mathcal{X}_i } \left| h_i (t_\nu,  x_i, \pi_i ) - x_i \right| \to 0 $ as $ k\to \infty $, meaning that for $ h_i $ convergence is uniform on compacts. 
Since $ f_i $ is Lipschitz continuous, it is uniformly continuous and bounded on compacts. Hence Lemma~1 of \cite{Lee2001} holds, and by a reasoning identical to the one above, if $ \mathcal{X} $ is a compact of $ \mathbb{R}^n $ we have:
 \[
 {\rm sup}_{x \in \mathcal{X} } \left| f_i ( h (t_\nu,  x, \pi )) - f_i(x) \right| \to 0 \qquad \text{as} \quad \nu \to \infty .
 \]
The argument holds independently for any component $ f_i $. 
Asymptotic time-independence and uniform convergence on compacts to $ f(x) $ follow consequently.
\qed

%---------------------------------------------------------------------------

\subsection{Proof of   Theorem~\ref{thm:glob-as-stab}}
By a standard converse theorem (e.g. Thm~4.14 of \cite{Khalil3rd}), global exponential stability of \eqref{eq:model} with $ f$ globally Lipschitz implies $ \exists $ a $ C^1 $ positive definite and radially unbounded Lyapunov function $ V \, :\, \mathbb{R}^n \to \mathbb{R}_+ $ and constants $ b_i >0 $, $ i=1, \ldots, 4$, such that $ \forall \, x \in \mathbb{R}^n $
\begin{align}
 b_1\| x\| ^2 & \leq V(x ) \leq b_2 \| x \|^2 \label{eq:gas_case_1}   \\
 \pde{V}{x} f(x) & \leq - b_3 \| x \| ^2  \label{eq:gas_case_2} \\
 \left\| \pde{V}{x} \right\| & \leq b_4 \| x \| . \label{eq:gas_case_3} 
\end{align}
With $ y = C( x + e^{-\Delta t} \gamma) $, the system \eqref{eq:model_xy} can be rewritten as 
\beq
\dot y =C\left( \dot x - \Delta e^{-\Delta t } \gamma \right) =  C f(y) - \begin{bmatrix} c_1 \delta_1 \gamma_1 e^{-\delta_1 t}  \\ \vdots \\ c_n \delta_n \gamma_n e^{-\delta_n t}  \end{bmatrix}.
\label{eq:doty}
\eeq
Considering $ V $ evaluated in $ y $, and computing its derivative along the trajectories of \eqref{eq:doty}, we get:
\beq
\begin{split}
\dot V & = \pde{V}{y} \dot y = 
\pde{V}{y} \left( \pde{h}{t} + \pde{h}{x} f(y) \right)_{x = h^{-1} (y) } \\
& = \pde{V}{y} C f(y) -  \pde{V}{y} C \Delta e^{-\Delta t } \gamma .
\end{split}
\label{eq:dotV-y}
\eeq
Defining $ k_1 = \| C \| >0 $, since \eqref{eq:gas_case_2} is valid everywhere, it is 
\[
\pde{V}{y} C f(y)  \leq  - k_1 b_3 \| y \| ^2  = - \alpha(\| y\|) ,
\]
for some $ \alpha\in \mathcal{K}_\infty^2 $, while the second term of \eqref{eq:dotV-y} can be rewritten as
\[
 \pde{V}{y} C \Delta e^{-\Delta t } \gamma  = \sum_i  \pde{V}{y_i} c_i  \delta_i \gamma_i e^{-\delta_i t } .
 \]
 For each $ t$, $ c_i \delta_i \gamma_i e^{-\delta_i t} \leq \zeta_1(t) \triangleq \max_i \left(  c_i   \delta_i | \gamma_i | \right) \max_i \left(   e^{-\delta_i t} \right)   \in \mathcal{L}^e$. 
Since 
\[
\pde{V}{y} = \pde{V}{x} \pde{x}{y} = \pde{V}{x} C^{-1} ,
\]
from \eqref{eq:gas_case_3}, Lemma~\ref{lemma:bound-affine-mask} (where we impose $ \sigma_i =0 $ and $ c_i = 1 + \phi_i $) and $ \| C^{-1} \| \leq 1 $, it is 
\[
\left\| \pde{V}{y} \right\| \leq b_4 \| x \|  \leq b_4 \| y \| + \zeta_2 (t) 
\]
for some $ \zeta_2 \in \mathcal{L}^e$. 
Hence, for some constant $ k_2>0$, 
\[
\begin{split}
\pde{V}{y} C  \Delta e^{-\Delta t } \gamma  & \leq k_2b_4 \| y \|  \zeta_1(t) + k_2 \zeta_1 (t) \zeta_2 (t) \\ 
& =  \beta(\| y \|, t) + \zeta_3 (t)
\end{split}
\]
with $ \beta \in \mathcal{KL}_\infty^{1,e} $ and $ \zeta_3 \in \mathcal{L}^e$.
Therefore, 
\[
\dot V \leq  - \alpha(\| y\|) + \beta(\| y \|, t) + \zeta_3 (t) 
\]
which has the same structure of \eqref{eq:compar_2}, meaning that we can apply Lemma~\ref{lemma:comparison2} and conclude that the system \eqref{eq:model} is uniformly globally attracted to $ x^\ast =0 $. Since \eqref{eq:glob-as-st-output-mask} is a privacy mask, $ \lim_{t\to\infty} y(t) =0$ and Assumption~\ref{ass1} holds, from Proposition~\ref{prop:ass1:indiscern}, \eqref{eq:model_xy} is a dynamically private version of \eqref{eq:model}.
\qed

%---------------------------------------------------------------------------

\subsection{Proof of   Corollary~\ref{cor:as-auton}}
Asymptotic autonomy of \eqref{eq:doty} is shown using an argument identical to that of the proof of Proposition~\ref{prop:no-equil}.
Convergence to the limit system $ \dot y = C f(y) $ and hence \eqref{eq:glob-as-st-limit} follows consequently. 
From expression \eqref{eq:doty} it is also clear that, for all $ y_o = h(0, x_o, \pi) $, $ \Omega_{y_o} = \{ 0 \} $, hence so it is for \eqref{eq:model_xy}.
\qed

%---------------------------------------------------------------------------

\subsection{Proof of   Theorem~\ref{thm:FJ}}
When $ \Theta \neq 0 $, $ -(L+\Theta) $ is Hurwitz, as can be easily deduced from e.g. \cite{Chen4232574}. 
In the $ z= x - x^\ast $ basis, for the unmasked system \eqref{eq:FJ_cont} a quadratic Lyapunov function can be used: $ V(z) = z^T P z $, where $ P=P^T>0$ is the solution of the Lyapunov equation 
\[
P(L+\Theta) + (L+\Theta)^T P = Q
\]
in correspondence of a given $ Q = Q^T >0$. 
The masked system \eqref{eq:FJ_cont_private} can be rewritten as
\beq
\begin{split}
\dot x  = & -(L + \Theta) \left(I + \Phi e^{-\Sigma t } \right)   x \\
& - L \left(I + \Phi e^{-\Sigma t } \right)  e^{-\Delta t} \gamma  \\
& + \Theta \left(I + \Phi e^{-\Sigma t } \right) x_o 
\end{split}
\label{eq:FJ_cont_proof1}
\eeq
or, in $ z$, after easy manipulations,
\[
\begin{split}
\dot z = &  (L+\Theta) \left(I + \Phi e^{-\Sigma t } \right)  z  \\
& - L  \left(I + \Phi e^{-\Sigma t } \right)  e^{-\Delta t} \gamma \\
& + \underbrace{ (L+ \Theta) \left[  (L+\Theta)^{-1} \Theta, \,  \left(I + \Phi e^{-\Sigma t } \right) \right]}_{\triangleq B(t) }  x_o
\end{split}
\]
where $ [ \,\cdot\,, \,\cdot \,] $ is the matrix commutator. Notice that for this term we have $ \| B(t) \| \leq \zeta_1(t) \in \mathcal{L}^e$. 
Inserting $ \dot z $ in $ \dot V $:
\beq
\begin{split}
\dot V = & - z ^T \Big( P(L+\Theta) \left(I + \Phi e^{-\Sigma t } \right) \\
&  \qquad + \left(I + \Phi e^{-\Sigma t } \right)  (L+\Theta)^T P \Big) z \\
& + 2 z^T \left( P (L+\Theta) B(t)  \right) x_o \\
& - 2 z^T PL  \left(I + \Phi e^{-\Sigma t } \right) e^{-\Delta t } \gamma .
\end{split}
\label{eq:FJ_cont_proof2}
\eeq
Looking at the terms of \eqref{eq:FJ_cont_proof2}:
\beqa
 - z ^T \Big( P(L+\Theta) \left(I + \Phi e^{-\Sigma t } \right) & & \nonumber \\
   + \left(I + \Phi e^{-\Sigma t } \right)  (L+\Theta)^T P \Big) z & \leq &- \alpha_1 (\| z \|), \label{eq:FJ_der_1}\\
 2 z^T \left( P (L+\Theta) B(t)  \right) x_o & \leq & \alpha_2 (\| z\|) \zeta_2 (t) , \label{eq:FJ_der_2}\\
 - 2 z^T PL  \left(I + \Phi e^{-\Sigma t } \right) e^{-\Delta t } \gamma & \leq & \alpha_3 (\| z\|) \zeta_3 (t) , \label{eq:FJ_der_3}
\eeqa
where $ \alpha_1 \in \mathcal{K}_\infty^2 $, $ \alpha_2, \, \alpha_3 \in \mathcal{K}_\infty^1$ and $ \zeta_i \in \mathcal{L}^e$, meaning that $ \beta_i (\| z \| , t) = \alpha_i (\| z \| ) \zeta_i (t) \in \mathcal{KL}_\infty^{1,e}$, $ i=2,\,3$. 
Therefore, overall we can write
\[
\dot V \leq - \alpha_1 (\| z \|) + \beta(\| z \|, t ) 
\]
where 
$
 \beta(\| z \|, t ) = \max_{j=2,3} \alpha_j (\| z \|) \max_{j=2,3} \zeta_j(t)  \in \mathcal{KL}_\infty^{1,e}.
 $
Since $ V$ is quadratic, positive definite, radially unbounded and vanishing in $ z=0 $, there exists two class $ \mathcal{K}_\infty^2$ functions $ \alpha_4 $ and $ \alpha_5 $ such that 
\beq
\alpha_4 (\| z \| ) \leq V(z) \leq \alpha_5 (\| z \| )  .
\label{eq:FJ_cont_proof3}
\eeq
Hence we can apply Lemma~\ref{lemma:comparison2} and obtain $ \lim_{t\to\infty} z(t) =0$. 
%the right hand side of \eqref{eq:FJ_cont_proof4} has the same structure as that of \eqref{eq:comparison1}. 
%Therefore we can apply the comparison lemma, using \eqref{eq:comparison1} with initial condition $ v(0) = V(0, z_o ) $.
%Denoting $ v(t) $ the corresponding solution, it is then 
%\beq
%V(t, z ) \leq v(t)\qquad  \forall \; t \geq 0 .
%\label{eq:FJ_cont_proof5}
%\eeq
%From Lemma~\ref{lemma:comparison1} and \eqref{eq:FJ_cont_proof5}, it follows that it must be $ \lim_{t\to\infty} V(t, z(t)) =0 $  for all $ z_o $, hence from \eqref{eq:FJ_cont_proof3} $ \lim_{t\to\infty} \alpha_4(\| z \| ) =0 $.
In the original variables $ x$, this implies $ \lim_{t\to\infty} x(t) =x^\ast (x_o)$ for all $ x_o$.
Convergence of $ x$ to $ x^\ast (x_o) $ is uniform in $ t$ because $ V$ does not depend on time.
\qed

%---------------------------------------------------------------------------

\subsection{Proof of   Corollary~\ref{cor:FJ-autonomous}}
The first part follows from Proposition~\ref{prop:no-equil} and the second from $ x^\ast(x_o) $ being a uniform attractor for each $ x_o $. 
\qed

%---------------------------------------------------------------------------

\subsection{Proof of   Theorem~\ref{thm:av-consensus}}
Notice first that the system \eqref{eq:consensus_xy} can be written as
\beq
\dot x = - L  \left(I + \Phi e^{-\Sigma t } \right)  \left( x + e^{-\Delta t} \gamma \right)  ,
\label{eq:consensus2}
\eeq
from which it is clear that the system \eqref{eq:consensus_xy} cannot have equilibrium points. 
It is also clear from \eqref{eq:consensus2} that $ \bfone^T \dot x =0 $ i.e., also \eqref{eq:consensus_xy} obeys to the conservation law $  \bfone^T  x(t) =  \bfone^T x_o = \eta \bfone $. 
As in the standard consensus problem \cite{Olfati2003Consensus}, we can therefore work on the $ n-1$ dimensional projection subspace $ {\rm span}(\bfone)^\perp $ and consider the time-varying Lyapunov function for the ``displacement vector'' $ x - \eta \bfone \in {\rm span}(\bfone)^\perp $: 
\[
V(t, x) = ( x - \eta \bfone )^T \left(I + \Phi e^{-\Sigma t } \right)  ( x - \eta \bfone ) .
\]
From now on we assume that all calculations are restricted to $ {\rm span}(\bfone)^\perp $.
The derivative of $ V$ along the solutions of \eqref{eq:consensus_xy} is 
\beq
\begin{split}
& \dot V (t, \, x ) =  \pde{V}{x} \dot x + \pde{V}{t} \\
 = &  -  2 ( x - \eta \bfone )^T   \left(I + \Phi e^{-\Sigma t } \right)  L  \left(I + \Phi e^{-\Sigma t } \right)  \left( x + e^{-\Delta t} \gamma \right)   \\
& - ( x - \eta \bfone )^T   \left( \Sigma \Phi e^{-\Sigma t } \right)  ( x - \eta \bfone ) \\
= &  - ( x - \eta \bfone )^T    \left(I + \Phi e^{-\Sigma t } \right)  (L + L^T )  \left(I + \Phi e^{-\Sigma t } \right) \left( x - \eta \bfone \right) \\
& -  \eta ( x - \eta \bfone )^T  \left(I + \Phi e^{-\Sigma t } \right)  ( L + L^T)  \left(I + \Phi e^{-\Sigma t } \right)  \bfone  \\
& -  ( x - \eta \bfone )^T  \left(I + \Phi e^{-\Sigma t } \right)  ( L + L^T)   \left(I + \Phi e^{-\Sigma t } \right)   e^{-\Delta t} \gamma  \\
& - ( x - \eta \bfone )^T   \left( \Sigma \Phi e^{-\Sigma t } \right)  ( x - \eta \bfone ) .
\end{split}
\label{eq:consensus2b}
\eeq
Since $ \phi_i >0$, it is $ 1 + \phi_i e^{-\sigma_i t } \geq 1 $ $ \forall \, t \geq 0$, and $ I + \Phi e^{-\Sigma t }  $ is a positive definite diagonal matrix, for the first term of \eqref{eq:consensus2b} we have
\[
\begin{split}
 & ( x - \eta \bfone )^T    \left(I + \Phi e^{-\Sigma t } \right)  (L + L^T )  \left(I + \Phi e^{-\Sigma t } \right) \left( x - \eta \bfone \right) \\
 & \geq 
  ( x - \eta \bfone )^T ( L + L^T) ( x - \eta \bfone ) \geq \alpha_1(\| x- \eta \bfone \| ) > 0 
 \end{split} 
 \]
 for some function $ \alpha_1 \in \mathcal{K}_\infty^2 $.
The second term of \eqref{eq:consensus2b} is linear in $ \| x- \eta \bfone \| $, and from $ L\bfone = L^T \bfone =0$, we have
 \[
\begin{split}
& - \eta ( x - \eta \bfone )^T  \left(I + \Phi e^{-\Sigma t } \right)  ( L + L^T)  \left(I + \Phi e^{-\Sigma t } \right)  \bfone \\
& = - \eta ( x - \eta \bfone )^T  \left(I + \Phi e^{-\Sigma t } \right)  ( L + L^T)   \Phi e^{-\Sigma t }   \bfone \\
& \leq \beta_1  (\| x- \eta \bfone \| , \, t) 
\end{split}
\]
for some function $ \beta_1 \in \mathcal{KL}_\infty^{1,e} $. Similarly, for the third term of \eqref{eq:consensus2b},
\[
\begin{split}
&  - ( x - \eta \bfone )^T  \left(I + \Phi e^{-\Sigma t } \right)  ( L + L^T)   \left(I + \Phi e^{-\Sigma t } \right)   e^{-\Delta t} \gamma  \\
& \quad  \leq  \beta_2  (\| x- \eta \bfone \| , \, t) 
\end{split}
\]
for some $ \beta_2 \in \mathcal{KL}_\infty^{1,e} $.
Finally, the fourth term of \eqref{eq:consensus2b} is
\[
( x - \eta \bfone )^T   \left( \Sigma \Phi e^{-\Sigma t } \right)  ( x - \eta \bfone ) = \alpha_2(\| x- \eta \bfone \| , \, t) 
\]
for some $ \alpha_2 \in \mathcal{KL}_\infty^{2,e} $, i.e., it is positive definite for all finite $t$, and vanishes as $ t\to\infty$. 
hence there exists a $ \alpha  \in \mathcal{K}_\infty^{2} $ such that 
\[
\alpha (v) \geq \alpha_1(v) + \alpha_2 (v, t) >0 \quad \forall \; v \in \mathbb{R}^+.
\]
Denote $
\beta  (\| x- \eta \bfone \| , t )   \in \mathcal{KL}_\infty^{1,e} $ a proper majorization of $ \beta_j (\| x- \eta \bfone \| , t ) $, $j=1, \, 2$.
Since, for all $ t$, $ V$ is quadratic, positive definite, radially unbounded and vanishing in $ x = \eta \bfone $, there exists two class $ \mathcal{K}_\infty^2$ functions $ \alpha_3 $ and $ \alpha_4 $ such that 
\beq
\alpha_3 (\| x- \eta \bfone \| ) \leq V(t, x) \leq \alpha_4 (\| x- \eta \bfone \| )  .
\label{eq:consensus3}
\eeq
Also in this case we can apply the comparison lemma, using \eqref{eq:comparison1} with initial condition $ v(0) = V(0, x_o ) $, where $ x_o $ such that $ \bfone^T x_o /n = \eta$.
%Denoting $ v(t) $ the corresponding solution, it is then 
%\beq
%V(t, x ) \leq v(t)\qquad  \forall \; t \geq 0 .
%\label{eq:consensus5}
%\eeq
From Lemma~\ref{lemma:comparison1}, % and \eqref{eq:consensus5}, 
it follows that it must be $ \lim_{t\to\infty} V(t, x(t)) =0 $  for all $ x_o $ such that $ \bfone^T x_o /n = \eta$, hence from \eqref{eq:consensus3} $ \lim_{t\to\infty} \alpha_3(\| x- \eta \bfone \| ) =0 $ or $ \lim_{t\to\infty} x(t) =\eta \bfone$ for all $ x_o$ such that $ \bfone^T x_o /n = \eta$.
%Since $ V$ is not time varying, convergence of $ x$ to $ \eta \bfone $ is uniform in $ t$. 
Since $ h(t, x, \pi)  = \left(I + \Phi e^{-\Sigma t } \right)  \left( x + e^{-\Delta t} \gamma \right)  $ is a privacy mask $ \lim_{t\to \infty} y(t) = \eta \bfone $ and Assumption~\ref{ass1} holds, from Proposition~\ref{prop:ass1:indiscern}, \eqref{eq:consensus_xy} is a dynamically private version of \eqref{eq:consensus1}.
\qed

%---------------------------------------------------------------------------

\subsection{Proof of   Corollary~\ref{cor:consensus-autonomous}}
Same as proof of Corollary~\ref{cor:FJ-autonomous}. \qed

%---------------------------------------------------------------------------

\subsection{Proof of   Theorem~\ref{thm:pinning}}
Notice first that \eqref{eq:glob-Lipsc-pinning} implies the following one-sided global Lipschitz condition used in \cite{Yu-doi:10.1137/100781699}:
\beq
(x-z)^T \big( f(x) - f(z) \big) \leq q (x-z)^T R (x-z) \quad \forall \; x, \, z \in \mathbb{R}^\nu.
\label{eq:onesided-glob-Lipsc-pinning}
\eeq
Denoting $ e_i (t) = x_i (t) - s(t) $ the error of the $ i$-th system from the desired trajectory, and using \eqref{eq:synchro_masked3}, then \eqref{eq:synchro_masked1} can be written in terms of $ e_i $ as
\[
\begin{split}
\dot e_i & = f(y_i) - f(s) 
- \sum_{j=1}^n \ell_{ij} R e_j\\
&- \sum_{j=1}^n \ell_{ij} R \Phi_j e^{-\Sigma_j t }  (e_j + s) \\
& - \sum_{j=1}^n \ell_{ij} R \left(I + \Phi_j e^{-\Sigma_j t } \right)   e^{-\Delta_j t } \gamma_j \\
&- p_i R e_i - p_i R \Phi_i e^{-\Sigma_i t }  (e_i + s) \\
&- p_i R \left(I + \Phi_i e^{-\Sigma_i t } \right) e^{-\Delta_i t} \gamma_i .
\end{split}
\]
Denote $ e= \begin{bmatrix} e_1^T & \ldots e_n^T\end{bmatrix}^T$ and, for brevity, $ \Psi_i(t) = I + \Phi_i e^{-\Sigma_i t }$.
A Lyapunov function, derived by that used in the standard pinned synchronization problem \cite{Yu-doi:10.1137/100781699}, is the following:
\[
V (e,t) = \sum_{i=1}^n e_i^T \Psi_i(t) \xi_i e_i.
\]
Since, for all $t$, $ V(t, e) $ is quadratic, positive definite, vanishing at $ e=0$, and radially unbounded, there exist two functions $ \alpha_1, \, \alpha_2 \in \mathcal{K}_\infty^2 $ such that 
\[
\alpha_1(\| e \| ) \leq V(t, e) \leq \alpha_2 (\| e \|) .
\]
For its derivative along the trajectories of \eqref{eq:synchro_masked1}-\eqref{eq:synchro_masked3} it is:
\begin{subequations}
\begin{align}
\dot V(t, e) & = \pde{V}{e} \dot e + \pde{V}{t}  \nonumber \ \\
& = 2 \sum_{i=1}^n e_i^T \Psi_i(t) \xi_i \dot e_i - \sum_{i=1}^n e_i^T \Big( \Sigma_i \Phi_i e^{-\Sigma_i t } \Big) \xi_i e_i   \nonumber \\
& =
2 \sum_{i=1}^n e_i^T \Psi_i(t) \xi_i \big( f(y_i) - f(s) \big)  \label{dotV_part1}\\
&
- 2 \sum_{i=1}^n e_i^T \Psi_i(t) \xi_i \sum_{j=1}^n \ell_{ij} R \Psi_j(t) e_j \label{dotV_part2} \\
&
- 2 \sum_{i=1}^n e_i^T \Psi_i(t) \xi_i \sum_{j=1}^n \ell_{ij} R \Phi_j e^{-\Sigma_j t }  s  \label{dotV_part4} \\
&
- 2 \sum_{i=1}^n e_i^T \Psi_i(t) \xi_i \sum_{j=1}^n \ell_{ij} R \Psi_j(t)   e^{-\Delta_j t } \gamma_j \label{dotV_part5} \\
&
-2 \sum_{i=1}^n e_i^T \Psi_i(t) \xi_i p_i R \Psi_i(t)  e_i  \label{dotV_part6} \\
&
- 2 \sum_{i=1}^n e_i^T \Psi_i(t) \xi_i p_i R  \Phi_i e^{-\Sigma_i t }  s  \label{dotV_part8} \\
&
- 2 \sum_{i=1}^n e_i^T \Psi_i(t) \xi_i p_i R \Psi_i(t)   e^{-\Delta_i t} \gamma_i \label{dotV_part9} \\
&
- \sum_{i=1}^n e_i^T \Big( \Sigma_i \Phi_i e^{-\Sigma_i t } \Big) \xi_i e_i . \label{dotV_part10} 
\end{align}
\end{subequations}
Of the eight terms on the right hand side, the first is the most complicated and will be treated last. Three other are quadratic in $ \| e \| $ and can be written as in \cite{Yu-doi:10.1137/100781699}, using Kronecker products:
\[
\begin{split}
 \eqref{dotV_part2}& % - \sum_{i=1}^n e_i^T \xi_i \sum_{j=1}^n \ell_{ij} R e_j 
= -  e^T \Psi(t) \Big( (  \Xi L + L^T \Xi )  \otimes R \Big) \Psi(t)  e \\
\eqref{dotV_part6} & %- \sum_{i=1}^n e_i^T \xi_i p_i R e_i 
=  -  2 e^T\Psi(t) \Big( \Xi P \otimes  R \Big) \Psi(t) e \\
 \eqref{dotV_part10} & = - e^T \Sigma \Phi e^{-\Sigma t } \Xi\otimes I   e 
\end{split}
\]
where $ \Psi = {\rm diag}(\Psi_1, \ldots, \Psi_n) $, $ \Sigma = {\rm diag}(\Sigma_1, \ldots, \Sigma_n) $ and $ \Phi = {\rm diag}(\Phi_1, \ldots, \Phi_n) $. 
The remaining four are all linear in $ \| e \| $ and decaying exponentially in $t$, and can be majorized in the following way
\[
\begin{split}
 \eqref{dotV_part4} & % \sum_{i=1}^n e_i^T \xi_i \sum_{j=1}^n \ell_{ij} R \Phi_j e^{-\Sigma_j t }  s \\
 \leq  k_1 \| e\| \! \cdot \! \| ( \Xi L+ L^T \Xi ) \otimes R \| \zeta_1 ( t) \\
& \leq \beta_1 (\| e\|, t ) 
\end{split}
\]
with $ \zeta_1  ( t) = \max_{i,j} \! \! \left\{ \left\| \Psi_i (0) \right\|  \!   \cdot  \!\left\| \Phi_j \right\| \!  \cdot  \! \|  s (t)  \|_\infty  \right\} \max_j \left\{ e^{-\Sigma_j T }\right\} \in \mathcal{L}^e$ ($s(t) $ is bounded for all $t$), $ k_1 >0$, and $ \beta_1 \in \mathcal{KL}_\infty^{1,e}$;
\[
\begin{split}
 \eqref{dotV_part5} & %  - \sum_{i=1}^n e_i^T \xi_i \sum_{j=1}^n \ell_{ij} R \left(I + \Phi_j e^{-\Sigma_j t } \right)   e^{-\Delta_j t } \gamma_j \\
 \leq  k_2 \| e\| \! \cdot \! \| ( \Xi L+ L^T \Xi ) \otimes R \|  \zeta_2 ( t) \\
& \leq \beta_2 (\| e\|, t ) 
\end{split}
\]
$ \zeta_2  ( t) = \max_j \left\{  \left\| \Psi_i(0) \right\| \!  \cdot  \!\left\| \Psi_j (0) \gamma_j \right\|  \right\}  \max_j \left\{ e^{-\Delta_j t } \right\}  \in \mathcal{L}^e$, $ k_2 >0$, and $ \beta_2 \in \mathcal{KL}_\infty^{1,e}$;
\[
\begin{split}
 \eqref{dotV_part8} & %  \sum_{i=1}^n e_i^T \xi_i p_i R  \Phi_i e^{-\Sigma_i t }  s \\
 \leq  k_3 \| e\| \! \cdot \! \|  \Xi P \otimes R \|  \zeta_1 ( t) \\
& \leq \beta_3 (\| e\|, t ) 
\end{split}
\]
$ k_3 >0$, $ \beta_3 \in \mathcal{KL}_\infty^{1,e}$;
\[
\begin{split}
 \eqref{dotV_part9} & % - \sum_{i=1}^n e_i^T \xi_i p_i R \left(I + \Phi_i e^{-\Sigma_i t } \right) e^{-\Delta_i t} \gamma_i \\
 \leq  k_4 \| e\| \! \cdot \! \|  \Xi P \otimes R \|  \zeta_2 ( t) \\
& \leq \beta_4 (\| e\|, t ) 
\end{split}
\]
$ k_4 >0$, $ \beta_4 \in \mathcal{KL}_\infty^{1,e}$.
Finally for \eqref{dotV_part1}, from \eqref{eq:inverse-mask-affine-II-1},
\[
\begin{split}
e_i & = x_i - s = \underbrace{ \left(I + \Phi_i e^{-\Sigma_i t } \right) ^{-1} }_{\triangleq F_i(t) } y_i - e^{-\Delta_i }\gamma_i - s \\ 
& = F_i(t) (y_i - s) + (F_i(t) - I ) s - e^{-\Delta_i }\gamma_i 
\end{split}
\]
where $ F_i(t) $ is diagonal, positive definite, $ \| F_i(t) \| \leq 1 $, and $ \lim_{t \to \infty} F_i(t) =I$.
Hence 
 \[
 \begin{split}
 \eqref{dotV_part1} & = 2 \sum_{i=1}^n \ (y_i - s)^T F_i(t)\Psi_i(t) \xi_i  \big( f(y_i) - f(s) \big)  \\
&   + 2\sum_{i=1}^n \left( s^T (F_i(t) -I)  - \gamma_i^T  e^{-\Delta_i t } \right) \Psi_i(t) \xi_i \big( f(y_i) - f(s) \big)  \\
& \leq 2 \sum_{i=1}^n q (y_i - s)^T \Psi_i(t) F_i(t) \Psi_i(t) \xi_i R (y_i -s )\\
 & \quad  +\beta_5(\| y_i- \bar s\|, \, t ) + \zeta_3 (t) 
 \end{split}
 \]
where for the first term we have used the one-sided Lipschitz condition
\[
\begin{split}
& (y_i - s )^T F_i(t) \Psi_i(t) \xi_i  \big( f(y_i) - f(s) \big) \\
& \quad \leq q (y_i-s)^T F_i(t) \Psi_i(t) \xi_i  R (y_i-s) 
\end{split}
\]
which follows from \eqref{eq:onesided-glob-Lipsc-pinning} and the equivalence of norms, and for the second term the fact that, from \eqref{eq:glob-Lipsc-pinning}, it depends linearly from $ \| y - \bar s \| $ and it decays exponentially to $0$ as $ t\to \infty $, meaning that $ \beta_5 \in \mathcal{KL}_\infty^{1,e}$ ($ \bar s $ is the vector of $n$ identical copies of $ s$). 
Furthermore, since, from Lemma~\ref{lemma:bound-affine-mask}, $ \| y - \bar s \| \leq k \| e \| + \zeta_4(t) $ for some $  \zeta_4 \in \mathcal{L}^e$ and $ k>1$, it is $  \beta_5(\| y-\bar s\|, \, t )  \leq \beta_6(\| e \| , \, t ) + \zeta_5(t) $ with $ \beta_6 \in \mathcal{KL}_\infty^{1,e} $ and $  \zeta_5 \in \mathcal{L}^e$. 
Inserting 
\[
y_i -s  =  \Psi_i(t) e_i + \Phi_i e^{-\Sigma_i t } s +  \Psi_i(t)   e^{-\Delta_i t } \gamma_i 
\]
and expanding, one gets a term quadratic in $ \| e \| $, 
\[
\sum_{i=1}^n q  e_i^T \Psi_i(t) \xi_i R \Psi_i(t)  e_i ,
\]
plus several other terms of first or zero order in $ \| e \| $, all vanishing exponentially fast in $t$. 
As long as $ s(t) $ is bounded, using arguments identical to those above, we can therefore write
 \[
 \eqref{dotV_part1} \leq  2 q e^T\Psi(t)  \Xi \otimes R\Psi(t)  e + \beta_7(\|e\|, t) + \zeta_6(t)
 \]
 with $ \beta_7 \in \mathcal{KL}_\infty^{1,e} $ and $ \zeta_6 \in \mathcal{L}^e$. 
 Putting together all terms quadratic in $ \| e \| $, since $ \Psi(t)  $ is diagonal positive definite and $ \Sigma \Phi e^{-\Sigma t} \Xi\otimes I $ is positive definite for all $ t$, it follows from \eqref{eq:synchro_ass2} that there exists $ \alpha_3 \in \mathcal{K}_\infty^2 $ such that 
 \[
 \begin{split}
&
 e^T \Psi(t) \left( 2\, q \Xi\otimes R - 2 \,\Xi P \otimes R - (\Xi L + L^T \Xi ) \otimes R \right) \Psi(t) e \\
 & - e^T \Sigma \Phi e^{-\Sigma t} \Xi\otimes I e 
% & \leq 
%   e^T \Psi(t) \left( 2\, q \Xi\otimes R - 2 \,\Xi P \otimes R - (\Xi L + L^T \Xi ) \otimes R \right) \Psi(t) e \\
 \leq  - \alpha_3 (\| e \|) .
 \end{split}
 \]
Hence 
\[
\dot V \leq  - \alpha_3 (\| e \|)  + \beta (\| e\|, t ) + \zeta(t)
\]
where $ \beta (\| e\|, t ) \in \mathcal{KL}_\infty^{1,e}  $ majorizes $ \beta_j  (\| e\|, t )  $, $ j =1, \ldots, 7$, and $ \zeta (t )  \in \mathcal{L}^e $ majorizes $  \zeta_j  ( t )  $, $ j=1, \ldots, 6 $, meaning that we can apply the comparison lemma (Lemma~\ref{lemma:comparison1}), using \eqref{eq:comparison1} with initial condition $ v(0) = V(0, e(0)) $, and the result follows. 
\qed

%---------------------------------------------------------------------------

\bibliographystyle{abbrv}

%
%{\small
% \bibliography{../../tex/bib/wireless,../../tex/bib/social,../../tex/bib/privacy,../../tex/bib/consensus,../../tex/bib/nonhol,../../tex/bib/linear}

\begin{thebibliography}{10}

\bibitem{Aeyels701102}
D.~Aeyels and J.~Peuteman.
\newblock A new asymptotic stability criterion for nonlinear time-variant
  differential equations.
\newblock {\em IEEE Transactions on Automatic Control}, 43(7):968--971, Jul
  1998.

\bibitem{Alaeddini7963642}
A.~Alaeddini, K.~Morgansen, and M.~Mesbahi.
\newblock Adaptive communication networks with privacy guarantees.
\newblock In {\em 2017 American Control Conference (ACC)}, pages 4460--4465,
  May 2017.

\bibitem{Altafini2019Dynamical}
C.~Altafini.
\newblock A dynamical approach to privacy preserving average consensus.
\newblock In {\em IEEE 58th Conf. on Decision and Control}, Dec 2019.

\bibitem{Ambrosin:2017:OOB:3155100.3137573}
M.~Ambrosin, P.~Braca, M.~Conti, and R.~Lazzeretti.
\newblock Odin: Obfuscation-based privacy-preserving consensus algorithm for
  decentralized information fusion in smart device networks.
\newblock {\em ACM Trans. Internet Technol.}, 18(1):6:1--6:22, Oct. 2017.

\bibitem{Arcak2001Observer}
M.~Arcak and P.~Kokotovic.
\newblock Observer-based control of systems with slope-restricted
  nonlinearities.
\newblock {\em IEEE Transactions on Automatic Control}, 46(7):1146--1150, Jul
  2001.

\bibitem{Artstein1976}
Z.~Artstein.
\newblock Limiting equations and stability of nonautonomous ordinary
  differential equations.
\newblock In J.~LaSalle, editor, {\em The stability of dynamical systems}, CBMS
  Regional Conference Series in Applied Mathematics. SIAM, Philadelphia, 1976.

\bibitem{ARTSTEIN1977184}
Z.~Artstein.
\newblock The limiting equations of nonautonomous ordinary differential
  equations.
\newblock {\em Journal of Differential Equations}, 25(2):184 -- 202, 1977.

\bibitem{Astolfi2007Remark}
A.~Astolfi.
\newblock A remark on an example by {T}eel \& {H}espanha with applications to
  cascaded systems.
\newblock {\em IEEE Transactions on Automatic Control}, 52(2):289--293, Feb
  2007.

\bibitem{CHAILLET2008519}
A.~Chaillet and D.~Angeli.
\newblock Integral input to state stable systems in cascade.
\newblock {\em Systems \& Control Letters}, 57(7):519 -- 527, 2008.

\bibitem{Chen4232574}
T.~Chen, X.~Liu, and W.~Lu.
\newblock Pinning complex networks by a single controller.
\newblock {\em IEEE Transactions on Circuits and Systems I: Regular Papers},
  54(6):1317--1326, June 2007.

\bibitem{Cortes7798915}
J.~Cort\'es, G.~E. Dullerud, S.~Han, J.~L. Ny, S.~Mitra, and G.~J. Pappas.
\newblock Differential privacy in control and network systems.
\newblock In {\em IEEE 55th Conf. on Decision and Control}, pages 4252--4272,
  Dec 2016.

\bibitem{Duan7402925}
X.~Duan, J.~He, P.~Cheng, Y.~Mo, and J.~Chen.
\newblock Privacy preserving maximum consensus.
\newblock In {\em 2015 54th IEEE Conference on Decision and Control (CDC)},
  pages 4517--4522, Dec 2015.

\bibitem{Dwork:2006:DP:2097282.2097284}
C.~Dwork.
\newblock Differential privacy.
\newblock In {\em Proceedings of the 33rd International Conference on Automata,
  Languages and Programming - Volume Part II}, ICALP'06, pages 1--12, Berlin,
  Heidelberg, 2006. Springer-Verlag.

\bibitem{Dwork:2014:AFD:2693052.2693053}
C.~Dwork and A.~Roth.
\newblock The algorithmic foundations of differential privacy.
\newblock {\em Found. Trends Theor. Comput. Sci.}, 9(3-4):211--407, Aug. 2014.

\bibitem{FAROKHI201713}
F.~Farokhi, I.~Shames, and N.~Batterham.
\newblock Secure and private control using semi-homomorphic encryption.
\newblock {\em Control Engineering Practice}, 67:13 -- 20, 2017.

\bibitem{FAROKHI2016254}
F.~Farokhi, I.~Shames, M.~G. Rabbat, and M.~Johansson.
\newblock On reconstructability of quadratic utility functions from the
  iterations in gradient methods.
\newblock {\em Automatica}, 66:254 -- 261, 2016.

\bibitem{FoAl2018}
A.~Fontan and C.~Altafini.
\newblock Multiequilibria analysis for a class of collective decision-making
  networked systems.
\newblock {\em Control of Networked Systems, IEEE Transactions on},
  5(4):1931--1940, 2018.

\bibitem{GUPTA20179515}
N.~Gupta, J.~Katz, and N.~Chopra.
\newblock Privacy in distributed average consensus.
\newblock {\em IFAC-PapersOnLine}, 50(1):9515 -- 9520, 2017.
\newblock 20th IFAC World Congress.

\bibitem{Hale8031339}
M.~T. Hale and M.~Egerstedt.
\newblock Cloud-enabled differentially private multi-agent optimization with
  constraints.
\newblock {\em IEEE Transactions on Control of Network Systems}, PP(99):1--1,
  2017.

\bibitem{He2019Consensus}
J.~{He}, L.~{Cai}, P.~{Cheng}, J.~{Pan}, and L.~{Shi}.
\newblock Consensus-based data-privacy preserving data aggregation.
\newblock {\em IEEE Transactions on Automatic Control}, pages 1--1, 2019.

\bibitem{Huang:2012:DPI:2381966.2381978}
Z.~Huang, S.~Mitra, and G.~Dullerud.
\newblock Differentially private iterative synchronous consensus.
\newblock In {\em Proceedings of the 2012 ACM Workshop on Privacy in the
  Electronic Society}, WPES '12, pages 81--90, New York, NY, USA, 2012. ACM.

\bibitem{Khalil3rd}
H.~Khalil.
\newblock {\em Nonlinear Systems}.
\newblock Pearson Education. Prentice Hall, 3rd edition, 2002.

\bibitem{Kia:RNC3178}
S.~S. Kia, J.~Cortes, and S.~Martinez.
\newblock Dynamic average consensus under limited control authority and privacy
  requirements.
\newblock {\em International Journal of Robust and Nonlinear Control},
  25(13):1941--1966, 2015.

\bibitem{Kogiso}
K.~{Kogiso} and T.~{Fujita}.
\newblock Cyber-security enhancement of networked control systems using
  homomorphic encryption.
\newblock In {\em 2015 54th IEEE Conference on Decision and Control (CDC)},
  pages 6836--6843, Dec 2015.

\bibitem{Lazzeretti6855039}
R.~Lazzeretti, S.~Horn, P.~Braca, and P.~Willett.
\newblock Secure multi-party consensus gossip algorithms.
\newblock In {\em 2014 IEEE International Conference on Acoustics, Speech and
  Signal Processing (ICASSP)}, pages 7406--7410, May 2014.

\bibitem{Lee2001}
T.-C. Lee, D.-C. Liaw, and B.-S. Chen.
\newblock A general invariance principle for nonlinear time-varying systems and
  its applications.
\newblock {\em IEEE Transactions on Automatic Control}, 46(12):1989--1993, Dec
  2001.

\bibitem{Liu2019Dynamical}
Y.~{Liu}, J.~{Wu}, I.~R. {Manchester}, and G.~{Shi}.
\newblock {Dynamical Privacy in Distributed Computing -- Part I: Privacy Loss
  and PPSC Mechanism}.
\newblock {\em arXiv e-prints}, page arXiv:1902.06966, Feb 2019.

\bibitem{Loria1393135}
A.~Loria, E.~Panteley, D.~Popovic, and A.~R. Teel.
\newblock A nested {M}atrosov theorem and persistency of excitation for uniform
  convergence in stable nonautonomous systems.
\newblock {\em IEEE Transactions on Automatic Control}, 50(2):183--198, Feb
  2005.

\bibitem{Manitara6669251}
N.~E. Manitara and C.~N. Hadjicostis.
\newblock Privacy-preserving asymptotic average consensus.
\newblock In {\em 2013 European Control Conference (ECC)}, pages 760--765, July
  2013.

\bibitem{Markus1956}
L.~Markus.
\newblock Asymptotically autonomous differential systems.
\newblock In S.~Lefschetz, editor, {\em Contribution to the theory of nonlinear
  oscillations}, Annals of Mathematical Studies. Princeton Univ. Press,
  Princeton, 1956.

\bibitem{Mo7465717}
Y.~Mo and R.~M. Murray.
\newblock Privacy preserving average consensus.
\newblock {\em IEEE Transactions on Automatic Control}, 62(2):753--765, Feb
  2017.

\bibitem{Monshizadeh19Plausible}
N.~Monshizadeh and P.~Tabuada.
\newblock Plausible deniability as a notion of privacy.
\newblock In {\em 59th IEEE Conference on Decision and Control (CDC)}, Dec
  2019.

\bibitem{MuASJC}
X.~Mu and D.~Cheng.
\newblock On the stability and stabilization of time-varying nonlinear control
  systems.
\newblock {\em Asian Journal of Control}, 7(3):244--255, 2005.

\bibitem{NOZARI2017221}
E.~Nozari, P.~Tallapragada, and J.~Cort\'es.
\newblock Differentially private average consensus: Obstructions, trade-offs,
  and optimal algorithm design.
\newblock {\em Automatica}, 81:221 -- 231, 2017.

\bibitem{LeNy6606817}
J.~L. Ny and G.~J. Pappas.
\newblock Differentially private filtering.
\newblock {\em IEEE Transactions on Automatic Control}, 59(2):341--354, Feb
  2014.

\bibitem{Olfati2003Consensus}
R.~Olfati-Saber and R.~Murray.
\newblock Consensus problems in networks of agents with switching topology and
  time-delays.
\newblock {\em Automatic Control, IEEE Transactions on}, 49(9):1520 -- 1533,
  sept. 2004.

\bibitem{PANTELEY1998Global}
E.~Panteley and A.~Loria.
\newblock On global uniform asymptotic stability of nonlinear time-varying
  systems in cascade.
\newblock {\em Systems \& Control Letters}, 33(2):131 -- 138, 1998.

\bibitem{PANTELEY2001Growth}
E.~Panteley and A.~Loria.
\newblock Growth rate conditions for uniform asymptotic stability of cascaded
  time-varying systems.
\newblock {\em Automatica}, 37(3):453 -- 460, 2001.

\bibitem{Pequito7039593}
S.~Pequito, S.~Kar, S.~Sundaram, and A.~P. Aguiar.
\newblock Design of communication networks for distributed computation with
  privacy guarantees.
\newblock In {\em 53rd IEEE Conference on Decision and Control}, pages
  1370--1376, Dec 2014.

\bibitem{PROSKURNIKOV201765}
A.~V. Proskurnikov and R.~Tempo.
\newblock A tutorial on modeling and analysis of dynamic social networks. part
  {I}.
\newblock {\em Annual Reviews in Control}, 43:65 -- 79, 2017.

\bibitem{Rezazadeh2018Privacy}
N.~Rezazadeh and S.~Kia.
\newblock Privacy preservation in a continuous-time static average consensus
  algorithm over directed graphs.
\newblock In {\em American Control Conference}, pages 5890--5895, 06 2018.

\bibitem{rouche2012stability}
N.~Rouche, P.~Habets, and M.~Laloy.
\newblock {\em Stability Theory by Liapunov's Direct Method}.
\newblock Applied Mathematical Sciences. Springer New York, 2012.

\bibitem{Ruan19}
M.~{Ruan}, H.~{Gao}, and Y.~{Wang}.
\newblock Secure and privacy-preserving consensus.
\newblock {\em IEEE Transactions on Automatic Control}, 64(10):4035--4049, Oct
  2019.

\bibitem{saberi1990global}
A.~Saberi, P.~Kokotovic, and H.~Sussmann.
\newblock Global stabilization of partially linear composite systems.
\newblock {\em SIAM Journal on Control and Optimization}, 28(6):1491--1503,
  1990.

\bibitem{Sontag2003RemarkCICS}
E.~D. Sontag.
\newblock A remark on the converging-input converging-state property.
\newblock {\em IEEE Transactions on Automatic Control}, 48(2):313--314, Feb
  2003.

\bibitem{Sontag2003Example}
E.~D. Sontag and M.~Krichman.
\newblock An example of a {GAS} system which can be destabilized by an
  integrable perturbation.
\newblock {\em IEEE Transactions on Automatic Control}, 48(6):1046--1049, June
  2003.

\bibitem{SONTAG1995351}
E.~D. Sontag and Y.~Wang.
\newblock On characterizations of the input-to-state stability property.
\newblock {\em Systems \& Control Letters}, 24(5):351 -- 359, 1995.

\bibitem{Sussmann1991Peaking}
H.~J. Sussmann and P.~V. Kokotovic.
\newblock The peaking phenomenon and the global stabilization of nonlinear
  systems.
\newblock {\em IEEE Transactions on Automatic Control}, 36(4):424--440, Apr
  1991.

\bibitem{Teel2004Examples}
A.~R. Teel and J.~Hespanha.
\newblock Examples of {GES} systems that can be driven to infinity by
  arbitrarily small additive decaying exponentials.
\newblock {\em IEEE Transactions on Automatic Control}, 49(8):1407--1410, Aug
  2004.

\bibitem{Wang7833044}
Y.~Wang, Z.~Huang, S.~Mitra, and G.~E. Dullerud.
\newblock Differential privacy in linear distributed control systems: Entropy
  minimizing mechanisms and performance tradeoffs.
\newblock {\em IEEE Transactions on Control of Network Systems}, 4(1):118--130,
  March 2017.

\bibitem{Wiese2016Secure}
M.~{Wiese}, K.~H. {Johansson}, T.~J. {Oechtering}, P.~{Papadimitratos},
  H.~{Sandberg}, and M.~{Skoglund}.
\newblock Secure estimation for unstable systems.
\newblock In {\em 2016 IEEE 55th Conference on Decision and Control (CDC)},
  pages 5059--5064, Dec 2016.

\bibitem{XUE2014852}
M.~Xue, W.~Wang, and S.~Roy.
\newblock Security concepts for the dynamics of autonomous vehicle networks.
\newblock {\em Automatica}, 50(3):852 -- 857, 2014.

\bibitem{Yu-doi:10.1137/100781699}
W.~Yu, G.~Chen, J.~L\"u, and J.~Kurths.
\newblock Synchronization via pinning control on general complex networks.
\newblock {\em SIAM Journal on Control and Optimization}, 51(2):1395--1416,
  2013.

\bibitem{ZHOU2008996}
J.~Zhou, J.~an~Lu, and J.~L\"u.
\newblock Pinning adaptive synchronization of a general complex dynamical
  network.
\newblock {\em Automatica}, 44(4):996 -- 1003, 2008.

\end{thebibliography}

%%%%%%%%%%%%%%%%%
\end{document}